\documentclass[12pt]{article}
\usepackage{amssymb,epsf,graphics,graphicx}

\catcode`\@=11
\@addtoreset{equation}{section}

\catcode`\@=11
\def\section{\@startsection{section}{1}{\z@}{3.5ex plus 1ex minus
   .2ex}{2.3ex plus .2ex}{\large\bf}}
\newcommand{\beq}{\begin{equation}}
\newcommand{\eeq}{\end{equation}}
\newcommand{\bea}{\begin{eqnarray*}}
\newcommand{\eea}{\end{eqnarray*}}
\newcommand{\beaq}{\begin{eqnarray}}
\newcommand{\eeaq}{\end{eqnarray}}
\def\ba{\begin{array}}
\def\ea{\end{array}}
\def\bi{\begin{itemize}}
\def\ei{\end{itemize}}
\def\bn{\begin{enumerate}}
\def\en{\end{enumerate}}

\begin{document}
\begin{flushright}
{\tt hep-th/0405162} \\
\end{flushright}
\vskip 1.5cm
\centerline{\Large \bf 
Boundary massive sine-Gordon model}
\centerline{\Large \bf
at the free Fermi limit and RG flow of Casimir energy}
\vskip 1cm
\centerline{\large Chaiho Rim} 
\vskip .5cm
\centerline{\it Department of Physics, 
Chonbuk National University}
\centerline{\it Chonju 561-756, Korea}
\centerline{\it email: rim@chonbuk.ac.kr}
\vskip 1cm
\vskip 2cm

\centerline{\bf Abstract}
RG flow of central charge $c_{\rm eff}$
is investigated  for the two boundary sine-Gordon model 
at the free Fermi limit. 
Thermodynamic Bethe ansatz approach is used to check the
non-monotonic decreasing properties of $c_{\rm eff}$, 
its resonance,
and the modification of $c_{\rm eff}$ 
due to the mismatch of the periodicity of 
a Lagrangian parameter 
and that of the boundary scattering  parameter. 
The detailed analysis uses the singularity structure 
of the system on the complex rapidity plane.

\section{Introduction} 
\noindent

The c-theorem states that effective central charge $c_{\rm eff}$
decreases monotonically as the bulk system size increases
\cite{c-theorem}.
This c-theorem is best studied and confirmed 
in the periodic bulk system of integrable field theories (IFT).
IFT provides a non-perturbative study of the scale 
dependence of  $c_{\rm eff}$ 
through thermodynamic Bethe ansats (TBA) \cite{TBA,YY}.
TBA is a set of integral equations 
for spectral densities of the system,
which uses the kernel 
derived from the scattering amplitude.
$c_{\rm eff}$ is written 
in terms of the weighted integration of particle energy.  
It is remarkable that IR properties 
obtained from scattering matrices
can be used to find the RG-flows from UV to IR. 
There are many examples 
which demonstrate the success of the TBA approach
\cite{TBA-RG1,TBA-RG2,TBA-RG3,TBA-RG4,TBA-RG5}. 

Notwithstanding the success,
$c_{\rm eff}$  does not obey c-theorem always.
In the presence of open  boundaries
non-monotonic behavior has been expected 
as well as resonances  \cite{bsG-GZ}.
In addition, in some cases  
TBA does not fully reflect the UV properties
of Lagrangian \cite{bLY,bsG-CSS1,bsG-R, bsG-CSS2}.
This originates from the singularity structure 
of the system. 

It is to be noted that 
even though the TBA equations are defined 
over the real rapidity variables,
the physical rapidity space 
is not restricted to the real rapidity line. 
For a periodic bulk system 
the rapidity is extended to complex values
with $-\pi < {\rm Im}(\theta) < \pi$.
The singlarity structure on the 
complex rapidity space provides 
all the information of particle spectra
including bound states.

If the system has a boundary, then 
the  physical rapidity plane reduces to 
$-\pi/2 < {\rm Im}\theta < \pi/2$
and RG flow of $c_{\rm eff}$ can be 
seen in terms of R-channel TBA \cite{R-TBA}. 
Depending on the boundary parameter ranges,
the RG flow of $c_{\rm eff}$ is 
known to be much affected 
by the existence of singularities 
near the real rapidity axis.
Especially, TBA may not have a finite solution
at a certain parameter value,
where singularities sit on the real rapidity.
In this case, TBA is not well-defined 
beyond the parameter range 
and should be modified
so that UV properties from the Lagrangian
matches with IR properties from the scattering matrix.
This demonstrates that 
complete understanding of the singularity 
structure is essential for understanding RG flow of 
$c_{\rm eff}$.
 
In this paper, we study the singularity structure 
of the boundary sine-Gordon model (bsG) on a strip.
bsG is given as   \cite{bsG-GZ} 
\bea
{\cal A}&=& \int_0^R dx \int_{-\infty}^\infty
\Bigg\{\frac1{4\pi}(\partial_{a}\varphi)^2
-2\mu(\cos(2 b \varphi)-1)\Bigg\}
\cr 
&& \qquad 
- 2\mu^{(L)}_{B}  \int_{-\infty}^\infty
\!\! dy \,
 \cos( b \phi_L-\chi_L)
-2\mu^{(R)}_{B}  \int_{-\infty}^\infty
\!\! dy \, 
\cos( b\phi_R-\chi_R )\,.
\eea
where $ \phi_L (y) = \varphi(0, y)$ 
( $ \phi_R (y) = \varphi(R, y)$)
is the boundary field
living at the left (right) edge.
$R$ is the width of the strip
representing the size of the system.
The coupling constant $b^2$ is real and 
is restricted to be less than 1.
($b^2 $ is scaled by $8\pi$
from the conventional choice 
$\beta^2 = 8\pi b^2$). 

In fact, the integrability allows 
additional boundary terms, 
\bea
- \frac{\alpha^{(L)} b}{\pi}\int_{-\infty}^\infty
\!\! dy \, \frac{d\, \phi_L(y)}{dy}
- \frac{\alpha^{(R)} b}{\pi}\int_{-\infty}^\infty
\!\! dy \, \frac{d\, \phi_R(y)}{dy}\,.
\eea
Unlike in the one-boundary action, one cannot eliminate
the whole term by introducing the bulk term 
\cite{bsG-AN}, 
\bea
- \frac{\alpha\, b}{\pi}
\int_0^R dx \int_{-\infty}^\infty
\!\! dy \, \partial_x \partial_y \varphi (x,y)\,.
\eea 
Therefore, the general action of bsG on a strip will be 
\beaq
{\cal A}&=& \int_0^R dx \int_{-\infty}^\infty
\Bigg\{\frac1{4\pi}(\partial_{a}\varphi)^2
-2\mu(\cos(2 b \varphi)-1)\Bigg\}
- \frac{\alpha\, b}{\pi}\int_{-\infty}^\infty
\!\! dy \, \frac{d\, \phi_R(y)}{dy}
\cr 
&& \qquad 
- 2\mu^{(L)}_{B}  \int_{-\infty}^\infty
\!\! dy \,
 \cos( b \phi_L-\chi^{(L)})
-2\mu^{(R)}_{B}  \int_{-\infty}^\infty
\!\! dy \, 
\cos( b\phi_R-\chi^{(R)})\,. 
\label{2bsG_action}
\eeaq
  
We will concentrate on the free fermionic
limit of bsG, 
$b^2 =1/2$ (or $\beta^2 = 4 \pi$)
\cite{bsG-free}.  
This limit is chosen for simplicity and clarity
since, even though (R-channel) TBA is known 
when the bulk scattering is diagonal,
there appears a complicated singularity structure 
and imposes instability in the numerical analysis
even in the massles case \cite{bsG-R,bsG-CSS2}.

The purpose of this paper is to understand  
RG-flow of $c_{\rm eff}$ and 
its dependence on the Lagrangian parameters 
through the singularity structure of the system.
In section \ref{sec:Massive}, we briefly 
summarize the relevant TBA analysis 
and analyze  the domain of parameters.
According to the boundary pole structure,  
the boundary parameters are classified into 
four Regimes: 
First, Lagrangian parameter $\chi$
is divided into two domains:
$0<\chi<\pi/2$ and its extended one
$\pi/2<\chi<\pi$.
Then, each domain is divided according to 
the scattering parameter $\eta$
appearing boundary scattering matrix 
in (\ref{bscattering2}) below:
$0<\eta<\pi/2$ and $\pi/2<\eta<\pi$.
In section \ref{sec:Massless}, 
massless limit of the theory is considered 
and the boundary parameter dependence 
of $c_{\rm eff}$ is summarized.  
Massless limit reduces the number 
of independent parameters
and simplify the analysis a little bit. 
In section \ref{sec:NN}, 
massive theory with small boundary strength 
$\mu_B$ at both edges (called NN-type)
is analyzed.
In  section \ref{sec:DD}, 
massive theory with large boundary strength 
$\mu_B$ at both edges (called DD-type) 
is analyzed.
In section \ref{sec:ND}, 
the case with small $\mu_B$ at one edge
and large $\mu_B$ the other edge
(called ND-type) is analyzed. 
Section \ref{sec:Conclusion} is the conclusion.
In Appendix Ising model is briefly presented. 
 
\section{Massive theory and boundary conditions}
\label{sec:Massive}
The TBA of the free Fermi limit is simply given as
\beq
\epsilon_0  = 2 r \cosh \theta
\eeq
with $M= 2 \pi \mu$ and $r=MR$.
This does not mean that the system is trivial.
All the essential difficulties 
lies in the non-trivial boundary action,
which induces complex singularity structures.
The effective central charge is given as 
\beq
c_{\rm eff}  
=  \frac{6r} {\pi^2 } \int_{-\infty}^\infty d \theta \,
\cosh \theta \, 
 \log Z(\theta) 
\equiv  c-24\Delta  \,,
\eeq 
where 
\beq
Z(\theta) = 1 + \lambda_0 (\theta) \, 
e^{- \epsilon_0}
+ \lambda_d (\theta) \, e^{-2 \epsilon_0}\,. 
\label{Z}
\eeq
Here $\lambda_0$ and $\lambda_d$ are fugacity 
due to the boundary effect, compactly written as
\cite{bsG-CSS1}
\bea
\lambda_0 (\theta)
&=&  \overline{K_L ^{++}} K_R^{++} 
+ \overline{K_L^{+-}} K_R^{+-} + \overline{K_L^{-+}} K_R^{-+} 
+\overline{K_L^{--}} K_R^{--} 
= \textrm{Tr}( \overline{K_L}  K_R )
\cr
\lambda_d (\theta)
&=& \overline{ ( K_L^{++} K_L^{--}
 -  K_L^{+-} K_L^{-+} ) }
(K_R ^{++} K_R ^{--} -  K_R ^{+-} K_R ^{-+}) 
= \textrm{Det} (\overline{ K_L} K_R )\,,
\eea
where the overlines denote for complex conjugate operation.
Subscript $L$ and $R$ stand for the left and right edge 
contribution. 
$K_{L,R}$ is  obtained from the 
boundary scattering amplitude 
$R_{L,R} ( \theta )$,
$K(\theta ) = R( i\frac\pi2 - \theta )$.
Explicitly they are written 
in terms of soliton and anti-soliton basis
\cite{bsG-GZ},
\beq
K^{\xi\, \xi} (\eta ,\vartheta |\theta) 
= P^{\xi} (\eta ,\vartheta |\theta )\,\,\,
G(\eta ,\vartheta | \theta) \,,
\quad
K^{\xi\,,-\xi} (\eta ,\vartheta  | \theta ) 
= Q^{\xi } (\eta ,\vartheta  | \theta)\,\,\,
G(\eta ,\vartheta  | \theta) 
\label{bscattering1}
\eeq
where $\xi=\pm$ and 
\beaq
P^{\xi }(\eta ,\vartheta  | \theta) 
& = & -i\sinh ( \theta )\cos (\eta )\cosh (\vartheta )
-\xi \cosh ( \theta )\sin (\eta )\sinh (\vartheta )
\cr 
Q^{\xi} (\alpha  | \theta) 
& = & i e^{- \xi 2 i \alpha} \sinh (\theta) \cosh (\theta )
\cr
G(\eta ,\vartheta  | \theta ) &=& 
\frac1{ 
\left( 2 \cos(\frac {\eta+i \theta}2)
\cos(\frac{\eta-i \theta}2) \right)\,
\left(2 \cosh(\frac {\vartheta+\theta}2)
\cosh(\frac{\vartheta- \theta}2) \right)} \,.
\label{bscattering2}
\eeaq
After some manipulation, $\lambda_d$ and $\lambda_0$ 
are written as
\beaq
\lambda_d &=&
\tan \Big(\frac{\eta_L -i \theta}2 \Big) \,
\tan \Big(\frac{\eta_L +i \theta}2 \Big) \,
\tanh \Big(\frac{\vartheta_L - \theta}2 \Big) \,
\tanh \Big(\frac{\vartheta_L + \theta}2 \Big) \,
\cr
&&
\tan \Big(\frac{\eta_R -i \theta}2 \Big) \,
\tan \Big(\frac{\eta_R +i \theta}2 \Big) \,
\tanh \Big(\frac{\vartheta_R - \theta}2 \Big) \,
\tanh \Big(\frac{\vartheta_R + \theta}2 \Big) \,
\label{lambdad}
\\
\lambda_0 &=& 2\Big( \sinh^2 \theta \, \cos \eta_L \,
\cos \eta_R\, \cosh \vartheta_L\, \cosh \vartheta_R 
+ \cosh^2 \theta\, \sin \eta_L\, \sin \eta_R\, 
\sinh\vartheta_L \sinh \vartheta_R 
\cr &&\quad
+ \cosh^2 \theta\, \sinh^2 \theta\, 
\cos (2\alpha) \Big)\,
G_L(\eta ,\vartheta ,\tilde u)\,
G_R(\eta ,\vartheta ,\tilde u)\,.
\label{lambda0}
\eeaq
From the fugacity
one can check the periodic property of $Z$ 
in (\ref{Z}),
\beq
Z(\theta+i 2\pi) = Z(\theta)\,,\qquad
Z(\theta+i\pi) 
= Z(\theta) \,
e^{2 \epsilon_0(\theta)}/\lambda_0(\theta)\,.
\eeq 

$K^{\xi\,,-\xi}$ in (\ref{bscattering1}) represents  the soliton number violation scattering 
at the edge
and contains the paremeter $\alpha$ which 
is identified with the Lagrangian parameter $\alpha$
in (\ref{2bsG_action}).
This is because the boundary derivative term
in (\ref{2bsG_action})
shifts the conjugate momentum of the field
$\Pi = \dot {\phi}/(2 \pi)$ 
by $- \alpha \delta(x-R)/\pi$,
and as the consequence,
the phase of the soliton operator in \cite{mandelstam}
is shifted by $\alpha$, $e^{i \alpha}$.

The scattering parameters $\eta$ and $\vartheta$ 
are related to the Lagrangian parameters 
$m_B \equiv  \mu_B \sqrt{2\pi/M}$
and $0 < \chi <\pi$ \cite{bsG-parameter}:
\beaq
&&\cos (  \eta/ 2)\, 
\cosh ( \vartheta /2 ) \, 
= m_B \, \cos  \chi 
\,, \qquad
\sin (  \eta /2)\, 
\sinh ( \vartheta/ 2 ) \, 
= m_B \, \sin  \chi  \,.
\label{bparameters}
\eeaq 
$\eta$ has the mirror symmetry with respect to 
$\chi=\pi/2$, $\eta(\chi)=\eta(\pi-\chi)$,
whose explicit relation is plotted in 
Figure~\ref{fig:eta_chi}.
\vskip 8pt
\[
\begin{array}{c}
\refstepcounter{figure}
\label{fig:eta_chi}
\epsfxsize=.50\linewidth
\epsfbox{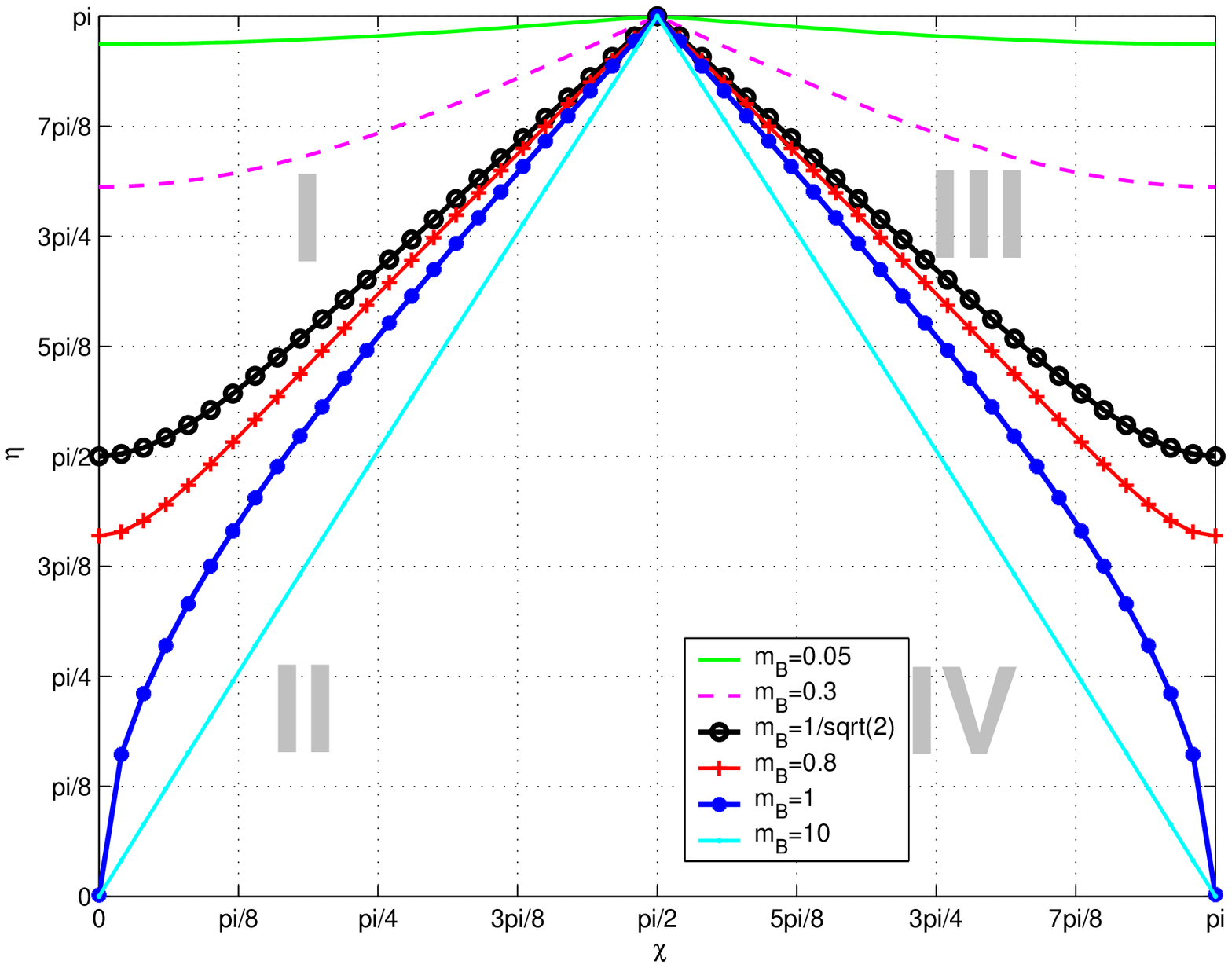}
\\
\parbox{.30\linewidth}{\small \raggedright
$\qquad$ Figure~\ref{fig:eta_chi}:%
~$\eta$ v.s. $\chi$}
\end{array}
\]

Note that the boundary bound state 
spectrum is \cite{bsG-GZ,b-state}
\beq
{\cal E}_B= M \cos \nu\,, \qquad 
\nu= \eta- \pi/2 \quad {\rm or}\quad  \pi- (\eta- \pi/2)
\eeq
where $0 <\nu <\pi/2$.
Thus, if $0< \eta < \pi/2$ 
there is no boundary bound state.
According to the existence of a boundary bound state,
we conveniently classify the boundary parameters into
four Regimes:
\vskip 0.3cm
\begin{tabbing}
\hskip 3cm
\= Case  $0< \chi  <\pi/2$ : \/ \= when  $\pi/2 < \eta < \pi$  \hskip 0.5cm  
\= Regime I  
\\ \> \> when $ 0 < \eta < \pi/2$ 
\> Regime II  \\ \\
\> Case  $\pi/2 < \chi  <\pi $ :\>
when $ \pi/2 < \eta < \pi$ \> Regime III \\
\>\> when $ 0 < \eta < \pi/2$ \> Regime IV.
\end{tabbing}
\vskip 0.3cm

It is convenient to define  $\chi_c$
which satisfies $\eta(\chi_c)= \pi/2$.
$\chi_c$ is restricted to the value  
$0 \le \chi_c \le \pi/2 $. 
When $0< m_B <1/\sqrt{2}$
(which will be called N-type boundary condition),
$\chi_c=0$ and only Regimes I and III are allowed.
On the other hand, when $ 1/\sqrt{2} < m_B $
(called D-type boundary condition),  
$0< \chi_c < \pi/4 $ 
and  all of the four Regimes are allowed.

One may wonder if Regime I and III (II and IV) 
are the same each other.
However, this is not the case. 
The reason is that scattering data 
and Lagrangian data 
do not coincide each other 
in Regime I and III (II and IV) : 
The periodicity of  $\eta_{L/R}$ 
in the scattering amplitude
differs from that in the Lagrangian.
$\eta_{L/R}$  in (\ref{bscattering2})
is periodic in $\pi$ and 
the symmetry $ \chi \to - \chi$
restricts the domain of $\chi$ as 
\beq
0 < \chi < \pi/2 \quad {\rm in}\;\,{\rm TBA}\,.
\eeq
On the other hand, 
$\chi$ in Lagrangian (\ref{2bsG_action})
is periodic in $2\pi$ 
and is restricted to 
\beq
 0 < \chi < \pi  \quad {\rm in}\;\,{\rm Lagrangian} \,.
\eeq
The periodicity of $\chi$ from the scattering data
does not match with the one given from the Lagrangian.
This suggests that $c_{\rm eff}$ 
obtained from TBA approach (\ref{ceff_0})
is in trouble when $\chi>\pi/2$ (Regime III and IV)
since TBA uses the IR scattering data. 

One would simply double the period of $\chi $ 
in (\ref{bscattering2}).
But this does not solve the problem.
Suppose one  extends  the  zone of  $\chi $ 
of one edge to $ 0 < \chi  <  \pi $. 
Then, $c_{\rm eff} (\chi) = c_{\rm eff} (\pi-\chi) $.
On the other hand, 
replacing $\chi$ into $\pi-\chi$ in the action 
will change the sign of the boundary term 
since $\mu_B $ will be replaced with $ - \mu_B$.
This means that 
$c_{\rm eff}$ is insensitive to 
the sign of the boundary action 
and  will have a cusp at $\chi=\pi/2$,
which  is not sound from 
the physics point of view \cite{bsG-CSS1}.
Thus $c_{\rm eff}$ given in (\ref{ceff_0})
has to be to modified in the extended parameter space. 
This possibilities are to be treated in detail 
in next sections.

The parameter $\alpha$ 
has no periodicity problem:
$\alpha$ in (\ref{bscattering2}) is periodic in $\pi$.  
On the other hand, $\alpha$-term
in the action (\ref{2bsG_action})
will give an exponential factor in the partition function 
$ e^{ - 2i \alpha 
\left( \frac{b\Delta \phi_0}{2\pi} \right) }$. 
Since $b \Delta \phi_0 /(2\pi) $ is an integer
(winding number),
the periodicity of $\alpha$ is $\pi$,
same as in the scattering data.
Thus, the range of $\alpha$ may be reduced to the domain   
$ -\pi/2 < \alpha < \pi/2\,$.
Using the symmetry $\alpha \to - \alpha$ in $\lambda_0$,
the domain can be further reduced to  
\beq
0< \alpha < \pi/2  \,.
\label{alpha-period}
\eeq

\section{Massless limit}
\label{sec:Massless}

The effect of the boundary parameters
on $c_{\rm eff}$ is easier to understand 
if the number of parameters can be reduced.  
The massless theory is a good example of this. 
In this section, we summarize  the singularity structure 
at the massless limit of bSG.

The massless limit is obtained 
by rescaling the parameters approriately
\cite{bsG-massless}.
Introducing a large rapidity parameter $\theta_0$, 
one rescales the mass scale
$ (M/2) e^{\theta_0} \to M $,
rapidity $\theta \to \theta + \theta_0$,
and boundary parameter 
$\vartheta \to \vartheta +  \theta_0  $.
Therefore, one has the parameter relation,
\beq
 e^{\vartheta } 
= 4 \pi   \mu_B^2 /M \,,
\qquad 
\eta= 2\chi\,.
\eeq
The rescaled quantities simplify the scattering data
of (\ref{bscattering1}) :  
\beq
K^{\xi \xi}(u) 
= -i \,e^{-\xi i \eta}  \, 
\frac {e^{\frac12( \vartheta -\theta)}} 
{2 \cosh \left(\frac{
\vartheta -\theta}{2} \right)} \,,\qquad
K ^{\xi, -\xi} (u) 
=i \,e^{-\xi 2i \alpha}  \, 
\frac {e^{-\frac12( \vartheta -\theta)}} 
{2 \cosh \left(\frac{
\vartheta -\theta}{2} \right)} \,.
\label{masslessK}
\eeq

The effective central charge is written as
\beq
 c_{\rm eff}  =  \frac{12r} {\pi^2 } 
\int_{-\infty}^\infty d \theta \,\,
 e^\theta \,\,   \log Z(\theta) 
\label{ceff_0}
\eeq 
where $Z(\theta)$ is the massless limit of (\ref{Z}) 
with  $\epsilon_0 =  2 r e^\theta $ and 
\beaq
\lambda_0 &=& 
\frac{
 \cos \eta\, 
e^{(\frac{\vartheta_L  +\vartheta_R }2 -  \theta)} 
+ \cos (2 \alpha) \,  
e^{-(\frac{\vartheta_L +\vartheta_R }2 - \theta)} } 
{2 \cosh (\frac{\vartheta_L  - \theta}2 )
\cosh (\frac{\vartheta_R  - \theta}2 )}
\cr
\lambda_d &=& \tanh \Big( \frac{\vartheta_L  - \theta}2 \Big) 
\tanh \Big( \frac{\vartheta_R  - \theta}2 \Big) \,.
\eeaq
In this massless limit,
only the combination of $\chi_L$ and $\chi_R$ 
enters in $c_{\rm eff}$ with 
$\eta=2\chi=2( \chi_R- \chi_L)$
and boundary bound state disappears
in (\ref{masslessK}).
This simplifies the analysis greatly.

The fugacity is  written in a simple form 
when both of the boundaries are given 
in an extreme condition so that 
they can be treated as a conformal field theory (CFT).
When both edges are in Dirichlet limit 
($\vartheta_L = \vartheta_R \to \infty$)
or in Neumann limit 
($\vartheta_L = \vartheta_R  \to -\infty$):
$ \lambda_0 =2 \cos \zeta$ and $\lambda_d = 1$
with $\zeta= 2\chi$ for Dirichlet 
and $\zeta= 2 \alpha$ for Neumann condition.
In this limit  
\beq
 c_{\rm eff} 
=  1-  {3 \zeta^2}/{\pi^2}\,,
\label{ceff_cft}
\eeq
where $ -\pi < \zeta <\pi$.
This result is obtained using 
the integration result for 
\bea
\int_0^\infty ds \log( 1 + 2 \cos \zeta \, e^{-s} + e^{-2s} ) 
= \frac{\pi^2}6 - \frac{\zeta^2}2 \,.
\eea
The  conformal dimension of the ground state is given as 
$\Delta = {\zeta^2}/( 8 \pi^2)$,
which reflects the duality 
between the role of $\alpha$ of Neumann condition
and the role of $\chi$ of  Dirichlet condition
in CFT of a compact boson theory.

When one edge is given in Neumann limit 
and the other edge in Dirichlet limit,
$ \lambda_0 = 0$ and $\lambda_d = -1$.
Then the effective central charge is independent of 
$\chi$ and $\alpha$, 
\beq
c_{\rm eff} = -\frac12\,,
\eeq
since
\bea
\int_0^\infty ds \log( 1 - e^{-2s} ) 
= -\frac{\pi^2}{12} \,.
\eea
The conformal dimension of the ground state is given as 
$\Delta =1/16$. 

The effect of $\alpha$ on $c_{\rm eff}$ 
is plotted in figure \ref{fig:ceff_alpha} at $\chi=0$. 
The dependence of $c_{\rm eff}$  on $\alpha$ 
is manifest when both boundaries are given as Neumann condition;
the smaller $\vartheta$ is, the greater the effect.
\vskip 8pt
\[
\begin{array}{c}
\refstepcounter{figure}
\label{fig:ceff_alpha}
\epsfxsize=.50\linewidth
\epsfbox{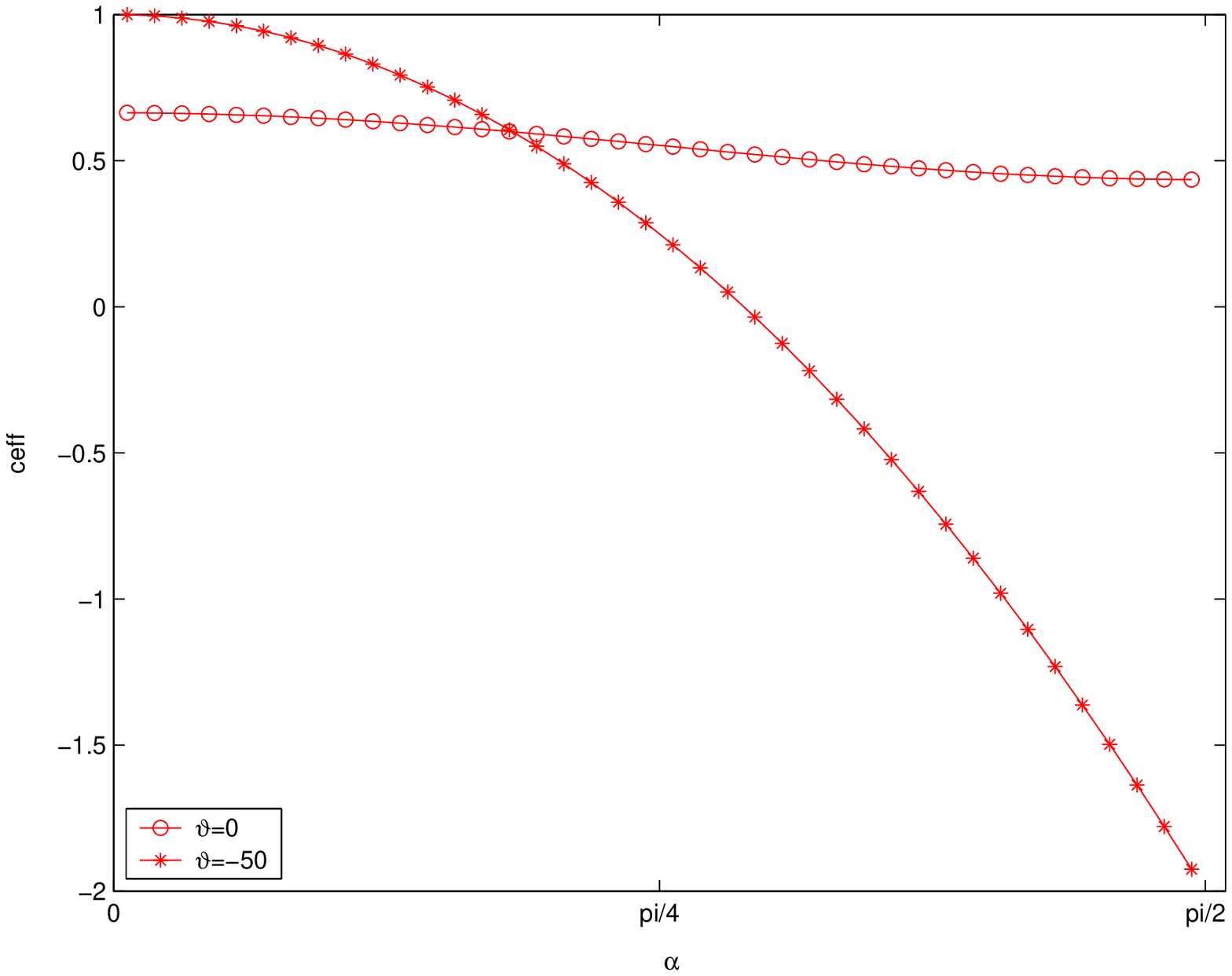}
\\
\parbox{.4\linewidth}{\small 
Figure \ref{fig:ceff_alpha}:~$c_{\rm eff}$~v.s.~$\alpha$
at $\chi_L=\chi_R=0$.}
\end{array}
\]
The figure \ref{fig:ceff_alpha} 
implies that
when plotting $c_{\rm eff}$~v.s.~$\vartheta$ 
with fixed $\alpha$ and $\chi$ 
 there will appear a bump 
at a sufficently large $\alpha$.
The bump is the result of the interference 
between the ground state
and an excited state.
This will be discussed in detail  
in section \ref{sec:NN}.

On the other hand, the branch-cut singularities arise
in $c_{\rm eff}$  when 
\beq
Z(\theta) =0\,.
\label{Z_0}
\eeq
This type of singularity will be called 
$Z_0$-singularity ($Z_0S$).
There are infinite number of $Z_0S$'s 
and all of them lie at Im($\theta)= \pm  \pi/2 $.
When $\eta=\pi$  ($\chi=\pi/2$) and $\alpha=0$, 
however, there appears  
$Z_0S$ at the real rapidity axis since 
\bea
Z(\theta) =\Bigg( 1 - \tanh \Big( 
\frac{\vartheta_L -\theta}2 \Big) \Bigg)
\Bigg( 1 - \tanh \Big( 
\frac{\vartheta_R -\theta}2 \Big) \Bigg)\,.
\eea
This is the source of trouble 
for the mismatch of the period of $\chi$ 
in two approaches, TBA and Lagrangian.
One cannot simply extend the domain of $\chi<\pi/2$ 
to $\chi>\pi/2$ 
but has to take care of the singularity 
at $\chi=\pi/2$  properly
\cite{bLY,bsG-CSS1,bsG-R,bsG-CSS2}. 

One way to proceed is to analytically 
continue $\chi<\pi/2$
by complexifying $\eta$ and rotating around at $\pi$
so that one can arrive at other domain of $\chi>\pi/2$.
Under this operation, one notices that 
for the ground state 
a  singularity 
at $ i \pi/2 + \theta_p$
($\theta_p$ is real and positive 
crosses the real rapidity line
and sits at $ -i \pi/2 + \theta_p$ 
(or vice versa) \cite{bsG-R}.
This singularity will be called 
$Z_0$-crossing singularity ($Z_0CS$).
 
$Z_0CS$ pushes the 
integration contour around the cross singularity 
position so that $c_{\rm eff}$ is put as 
\beq
c_{\rm eff} = -\frac {24r}{\pi} \,e^{\theta_p}
+\frac{12r}{\pi^2} -\!\!\!\!\!\!\int _{-\infty}^\infty 
d\theta \,e^\theta \, 
\log Z(\theta) \,.
\label{ceff_0ext}
\eeq
The first term in RHS 
is due to the singularity crossing 
and the second one is the real axis contribution
where $ -\!\!\!\!\!\int $ represents 
the principal value of the integration.

Figure \ref{fig:ceff_chi} 
is  the plot obtained from (\ref{ceff_0})
when  $\chi< \pi/2$,
and  (\ref{ceff_0ext}) when  $\chi>\pi/2$.
In this modified version, the curve is  not cusped  
at $\chi=\pi/2$, 
but is smooth and the period of
$c_{\rm eff}$ matches 
with the period of Lagrangian.
\vskip 8pt
\[
\begin{array}{c}
\refstepcounter{figure}
\label{fig:ceff_chi}
\epsfxsize=.50\linewidth
\epsfbox{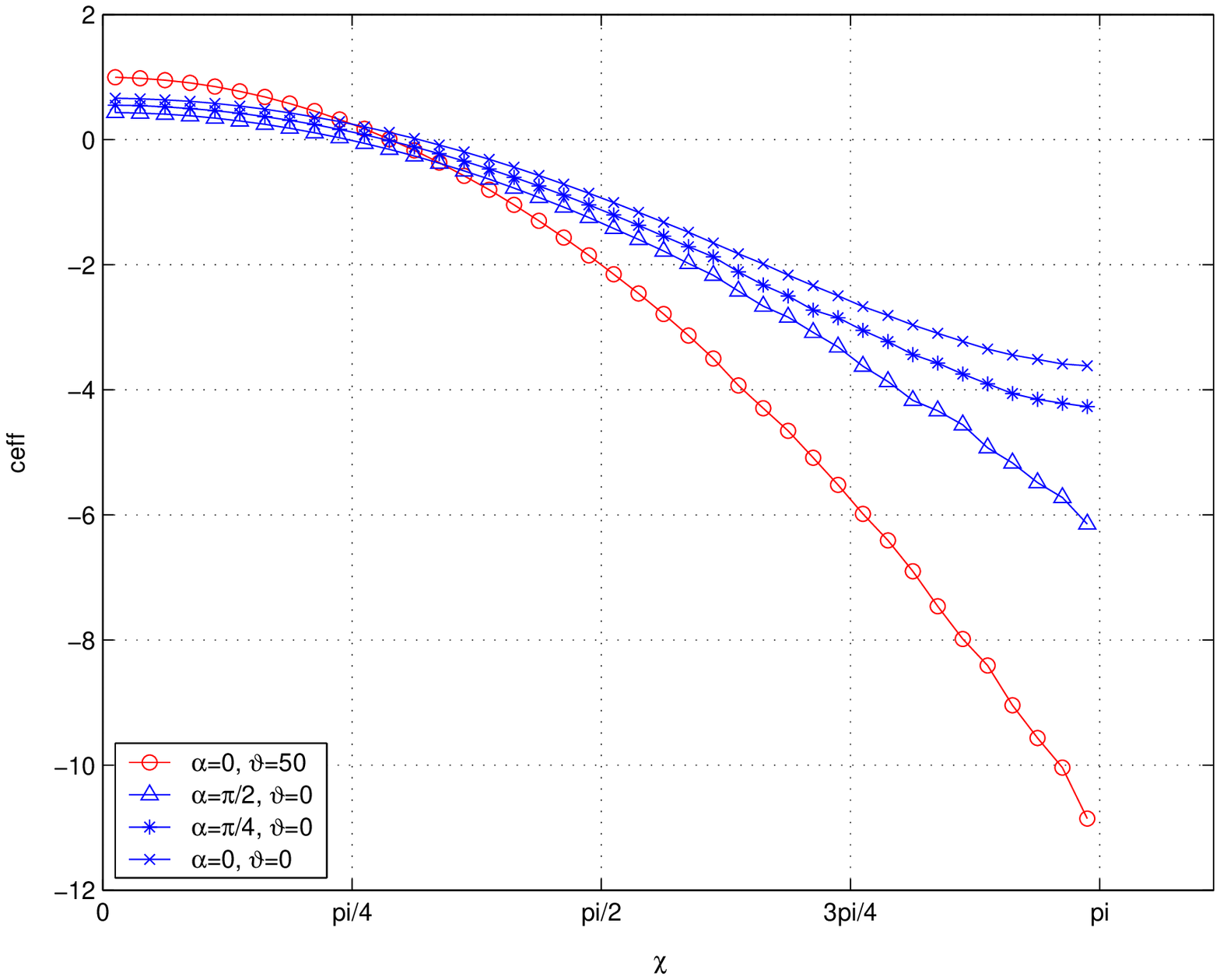}
\\
\parbox{.6\linewidth}{\small 
Figure \ref{fig:ceff_chi}:
~$c_{\rm eff}$~v.s.~$\chi$ for $0 < \chi <\pi$
from the modified $c_{\rm eff}$.}
\end{array}
\]

\section{NN-type boundary condition}
\label{sec:NN}
In this section, the boundary strength
$m_B$ of both edges are considered to be 
small, $m_B^{(L,R)} < 1/\sqrt2\,$ (called NN-type). 
In this case the boundary bound state exists
and  Regime I and III ($\pi/2<\eta< \pi$) are allowed.

\vskip 8pt
\[
\begin{array}{c}
\refstepcounter{figure}
\label{fig:ceff_m0_2}
\epsfxsize=.45\linewidth
\epsfbox{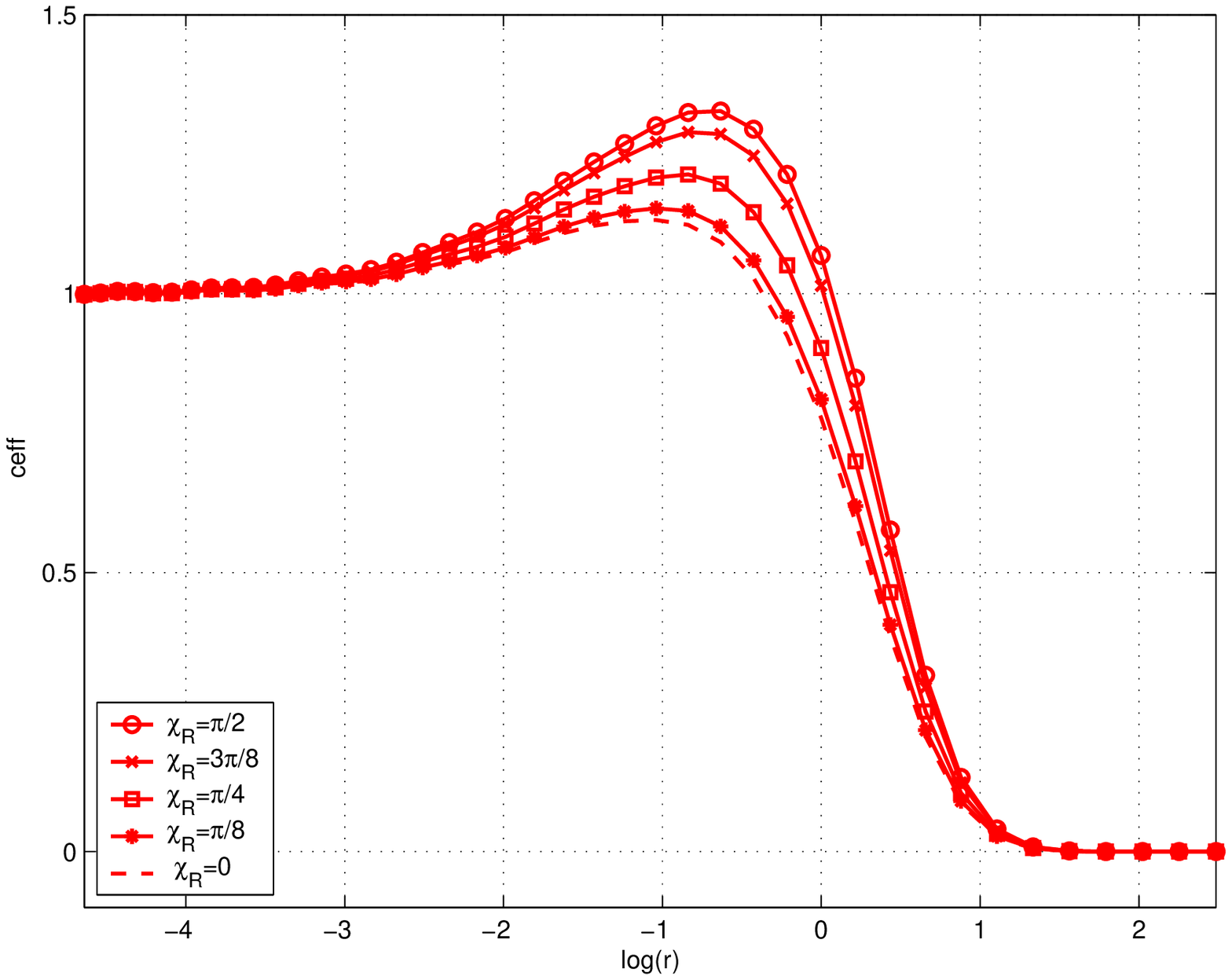}
\qquad
\refstepcounter{figure}
\label{fig:ceff_n}
\epsfxsize=.45\linewidth
\epsfbox{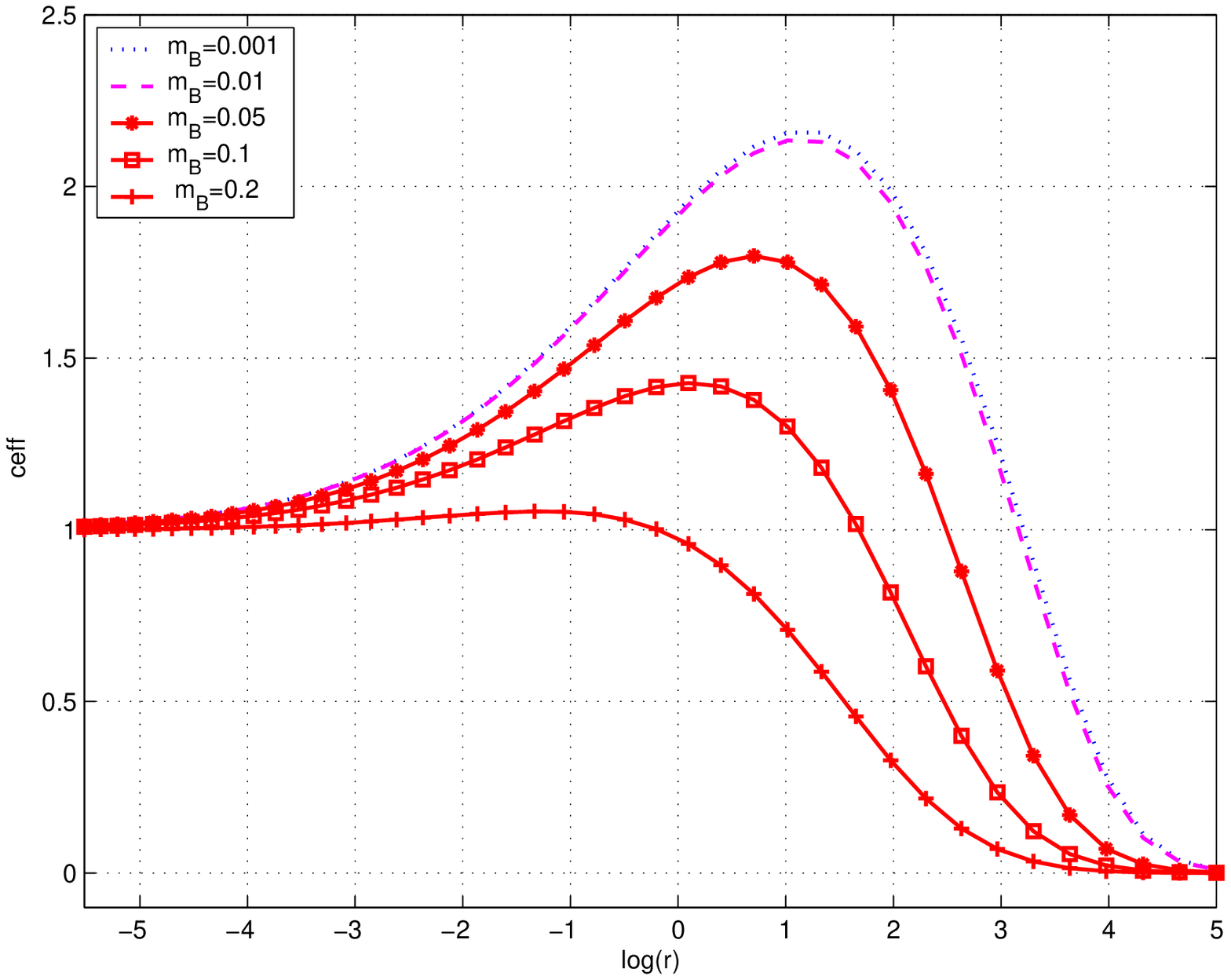}
\\
\quad
\parbox{.45\linewidth}{\small 
Figure~\ref{fig:ceff_m0_2}:%
~$c_{\rm eff}$~v.s.~$\log(r)$ 
at $m_B^{(L)}=m_B^{(R)} =0.2$ and
$\chi_L=\alpha=0$.}
\qquad
\parbox{.45\linewidth}{\small 
Figure~\ref{fig:ceff_n}:%
~$c_{\rm eff}$~v.s.~$\log(r)$ at 
$\chi_R=\chi_L=0$ and $\alpha=0$.
Here $m_B= m_B^{(L)}=m_B^{(R)} $.}
\end{array}
\]
\vskip 8pt

In figure \ref{fig:ceff_m0_2}, 
$c_{\rm eff} $ is plotted 
against $\log(r)$ for Regime I 
when $\alpha=0$. 
In UV limit ($r \to 0$), $c_{\rm eff} \to1$
and is independent of $\chi$.
This limit corresponds to the Neumann limit of the 
massless theory.
In IR limit ($r \gg 1$), $c_{\rm eff} \to 0$.
In between UV and IR, 
$c_{\rm eff}$ is not monotonically 
decreasing but there appears a resonance bump.
This resonant bump is enhanced as $m_B^{(L/R)}$ approaches 0,
which is seen in figure~\ref{fig:ceff_n}.

The resonance implies that the ground state 
interferes with the excited state,
the boundary bound state.
This is explained from the singularity structure.
Note that a boundary bound state provides 
a pair of poles in $Z$ (\ref{Z}),
which will be called $Z_\infty$ singularity 
($Z_\infty S$),
\beq
Z=\infty \,.
\label{Z_i}
\eeq
$Z_\infty S$ lies at $ \pm i  (\pi - \eta)$ 
with $0< (\pi - \eta)  <\pi/2$.
This is shown in figure~\ref{fig:Zi_contour_theta_n}.
\vskip 8pt
\[
\begin{array}{c}
\refstepcounter{figure}
\label{fig:Zi_contour_theta_n}
\epsfxsize=.45\linewidth
\epsfbox{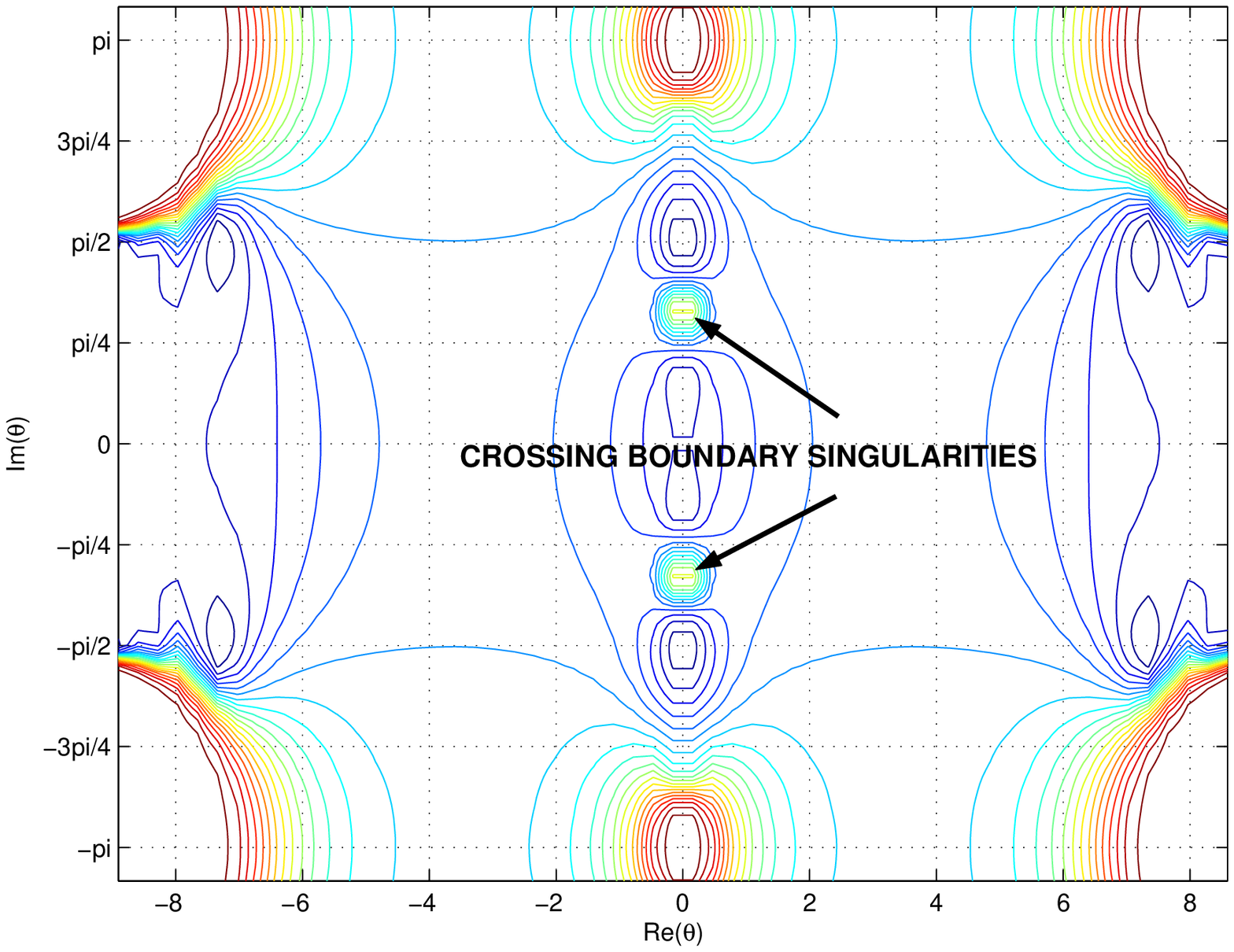}
\qquad
\refstepcounter{figure}
\label{fig:Z_contour_theta_n}
\epsfxsize=.45\linewidth
\epsfbox{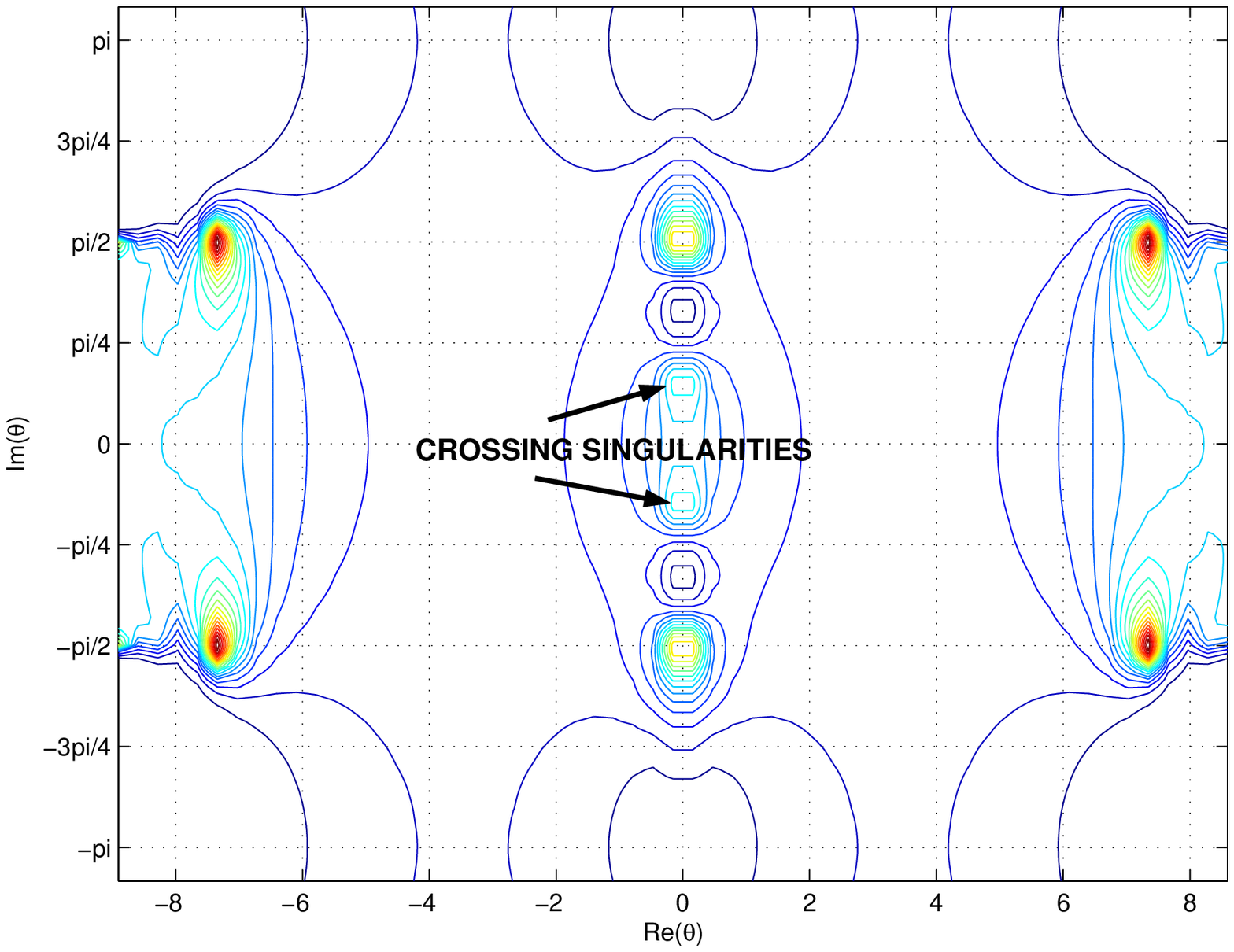}
\\
\qquad
\parbox{.45\linewidth}{\small 
Figure \ref{fig:Zi_contour_theta_n}:%
~Contour plot of $Z_\infty$ singularities.
Crossing singularities
lie at the imaginary axis 
in the physical strip.}
\qquad
\parbox{.45\linewidth}{\small \raggedright
Figure \ref{fig:Z_contour_theta_n}:%
~Contour plot of $Z_0$ singularities.
Crossing singularities are squeezed between
$Z_\infty$ crossing singularities.}
\end{array}
\]
\vskip 8pt

Most of $Z_0S$'s lie at $ \pm (i \pi/2 \pm \theta_D )$,
where $\theta_D>0$ is real
and depends on the scale as well as 
the boundary parameters.
At $\theta_D=0$, there is a ubiquitous 
singularity pair.
In addition to this, 
an additional pair exists at $\pm i \theta_N$ 
in this NN-type boundary condition
where $0< \theta_N <(\pi -\eta)$.
As $r \gg 1$, $\theta_N \to (\pi - \eta) $.
The singularities are 
given in figure~\ref{fig:Z_contour_theta_n}.

In this NN-type, 
the ground state with $m_B$ can be degenerate with
the ground state with $-m_B$
as $r \to \infty $ and $m_B \to 0$.  
In the intermediate scale, the two states
are not completeley degenerate 
but interfere with each other.
This is because as $\chi \to \pi/2$ or $m_B \to 0$,  
$\eta \to \pi$ and  
$Z_\infty CS$ at $\pm i (\pi -\eta)$
moves close to the real rapidity line. 
At the same time $Z_0 CS$ sqeezed 
between this $Z_\infty CS$ 
moves close to the real line too.
In fact, $Z_\infty S$ at $\pm i (\pi -\eta)$
and $Z_0S$ at $\pm i \theta_N$ are crossing
and the ground state with $-\mu_B$ 
can be obtained from the ground state with $\mu_B$ 
by taking care of the crossing singularity.
(See Eq.~(\ref{ceff_NN_ext})  below). 
As the singularities lie near the real rapidity axis
the two states interfere each other strongly.
This resonance effect is first noticed in Ising system 
in the presence of weak magnetic field in \cite{bsG-GZ}.
(See Appendix for Ising model). 

In Regime III ($\chi>\pi/2$), 
one has to count in
crossing singularities,
which cure $c_{\rm eff}$ to have the right 
periodicity of $\chi$. 
When  $\chi=\pi/2$, $\eta=\pi$
and the singularity sits at $\theta=0$.
Thus $\chi=\pi/2$ is the 
singular point in the parametric space.
One may analytically continue $\chi$ 
into the extended domain $\pi/2<\chi<\pi$.
(Equivalently, one may analytically continue 
$m_B$ around $m_B=0$.
See this possibility in 
boundary Lee-Yang case \cite{bLY})

Under the analytic continuation, 
$Z_\infty CS$  as well as $Z_0CS$ 
is crossing at the same time.
The ground state contribution is obtained by  
minimizing the number of crossing 
singularities and 
taking the crossing singularities  
close to the imaginary axis.

In figure~\ref{fig:Zi_contour_theta_n}
and \ref{fig:Z_contour_theta_n},
the the crossing singularities 
are pointed with arrows.
$Z_\infty CS$ is given as $\pm i (\pi-\eta_R)$ 
and $Z_0CS$ as $ \pm i \theta_N$.
When only one edge's $\chi$ exceeds $\pi/2$,
$c_{\rm eff}$ is modified as
\beq
c_{\rm eff} 
= -\frac {24r}{\pi} \,
\Big(\sin {\theta_N} - \sin \eta_R \Big)
+\frac{6r}{\pi^2} -\!\!\!\!\!\!\int _{-\infty}^\infty d \theta
\,e^\theta \, 
\log Z(\theta)  \,,
\label{ceff_NN_ext}
\eeq
where the right edge is chosen 
to have $\chi_R>\pi/2$.
$\theta_N$ is the real function of 
$r$, $m_B$, $\chi$, and $\alpha$\/.
Even though the exact dependence 
of $\theta_N$ is not known analytically,
it can be calculated numerically.
When $r \to \infty$, $\theta_N \to \eta_R$.
When $r \to 0$, $\theta_N \to \theta_0$
and $\theta_0$ vanishes as $m_B \to 0$.
At UV limit, $c_{\rm eff} =1$
and at IR limit,  $c_{\rm eff} =0$.
In figure~\ref{fig:ceff_m0_2_ext}, plotted is 
$c_{\rm eff}$~v.s.~$\log(r)$ when  $m_B^{(L/R)}=0.2$,
and $\alpha=\chi_L=0$.
\vskip 8pt
\[
\begin{array}{c}
\refstepcounter{figure}
\label{fig:ceff_m0_2_ext}
\epsfxsize=.60\linewidth
\epsfbox{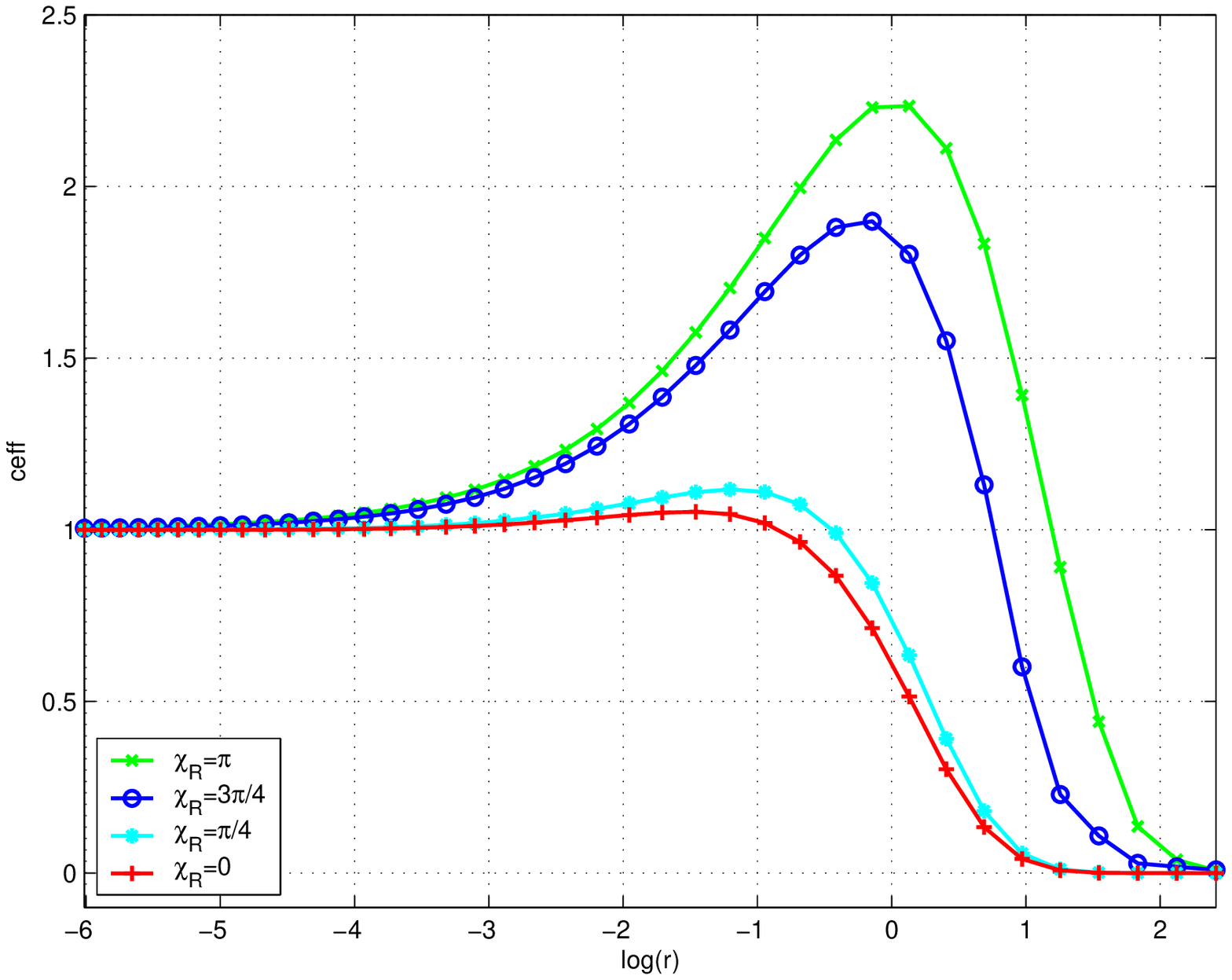}
\\
\parbox{.95\linewidth}{\small \raggedright
Figure~\ref{fig:ceff_m0_2_ext}:%
~$c_{\rm eff}$~v.s.~$\log(r)$ 
at $m_B^{(L)}= m_B^{(R)}= 0.2$ and
$\alpha=\chi_L=0$. Curves with 
$\chi_R=0, \pi/4$ belong to Regime I,
and $\chi_R=3\pi/4, \pi$ 
to Regime III.}
\end{array}
\]
\vskip 8pt

In figure \ref{fig:ceff_m0_2_ext}
$m_B^{(L)}=m_B^{(R)}$ is used.
When $m_B^{(L)} \ne m_B^{(R)}$
one needs to calculate $c_{\rm eff}$ in two ways 
and chooses the minimum 
as the ground state result among the two :
One with $(\chi_L, \chi_R)$ 
and the other with 
$(\widetilde {\chi_L}, \widetilde {\chi_R})$ 
where 
$\widetilde {\chi_L}= \pi -\chi_L$ and  
$\widetilde {\chi_R}=\pi -\chi_R$.
This is because bsG with $(\chi_L, \chi_R)$ 
is equivalent to one with 
$(\widetilde {\chi_L}, \widetilde {\chi_R})$ 
under the field operation,
$b\,\varphi \to b\,\varphi +\pi$ and
$(\varphi \to - \varphi)$,
but the crossing singularity contribution can 
be different.

Suppose both $\chi$'s exceed $\pi/2$,
(e.g., $\chi_L> \pi/2$ and $\chi_R> \pi/2$).
Then one may rearrange both parameters as 
$\widetilde{\chi} < \pi/2$
so that the crossing singularities 
do not contribute. 
If one would count in  
the crossing singularities of both edges,
then $c_{\rm eff}$ would 
be of an excited state result 
rather than of the ground state one. 

The parameter $\alpha$ also affects  $c_{\rm eff}$.
Especially, at UV limit the effect is manifest 
and $c_{\rm eff} \ne 1$ when $\alpha\ne 0$. 
One may convince oneself  that at UV limit 
$c_{\rm eff}$~v.s.~$\alpha$ reproduces the CFT limit
(\ref{ceff_cft}).
This behavior is plotted in figure \ref{fig:ceff_n_alpha}.
(Refer to figure \ref{fig:ceff_alpha_m100} below also).
\vskip 8pt
\[
\begin{array}{c}
\refstepcounter{figure}
\label{fig:ceff_n_alpha}
\epsfxsize=.60\linewidth
\epsfbox{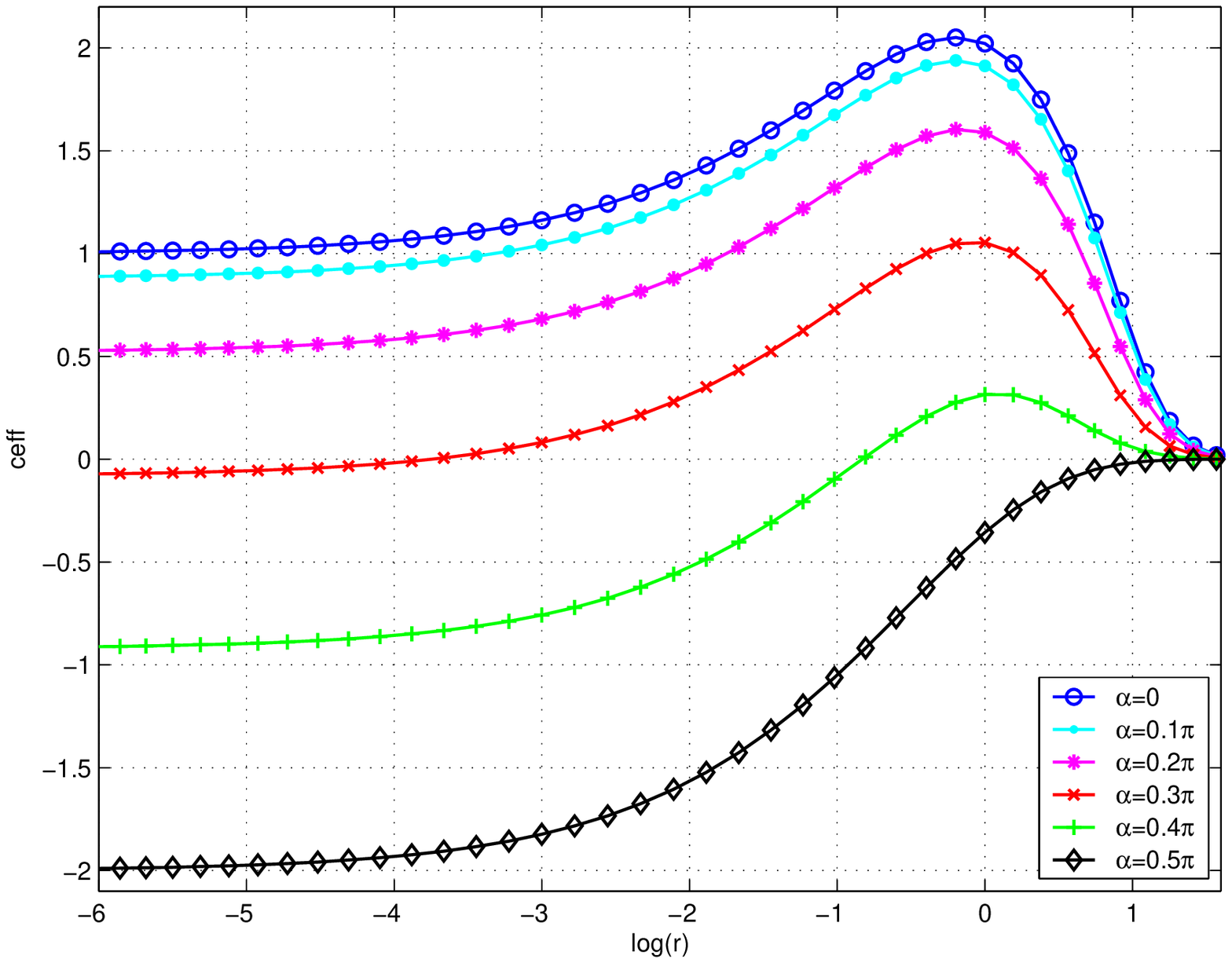}
\\
\parbox{.85\linewidth}{\small 
Figure~\ref{fig:ceff_n_alpha}:%
~$c_{\rm eff}$~v.s.~$\log(r)$ at 
$\chi_R=\chi_L=0$ and  
$m_B= 0.001 $
with  various $\alpha$.}
\end{array}
\]
\vskip 8pt
At IR limit, however, $\alpha$ effect vanishes 
and $c_{\rm eff} \to 0$. 
Therefore, RG-flow between UV and IR is not monotonically 
decreasing. 
A resonance bump is also seen 
at the intermediate scale due to 
the singularities near the real axis.
\vskip 0.5cm 

\section{DD-type boundary condition}
\label{sec:DD}

In this section, both edges are 
restricted to DD-type 
($ m_B^{(L/R)} >1/\sqrt{2}$).
In DD-type, 
all the four Regimes are to be considered.
To simplify the analysis, we put 
$m_B =m_B^{(L)}=m_B^{(R)}$. 

\[
\begin{array}{c}
\refstepcounter{figure}
\label{fig:ceff_logr_m1000}
\epsfxsize=.50\linewidth
\epsfbox{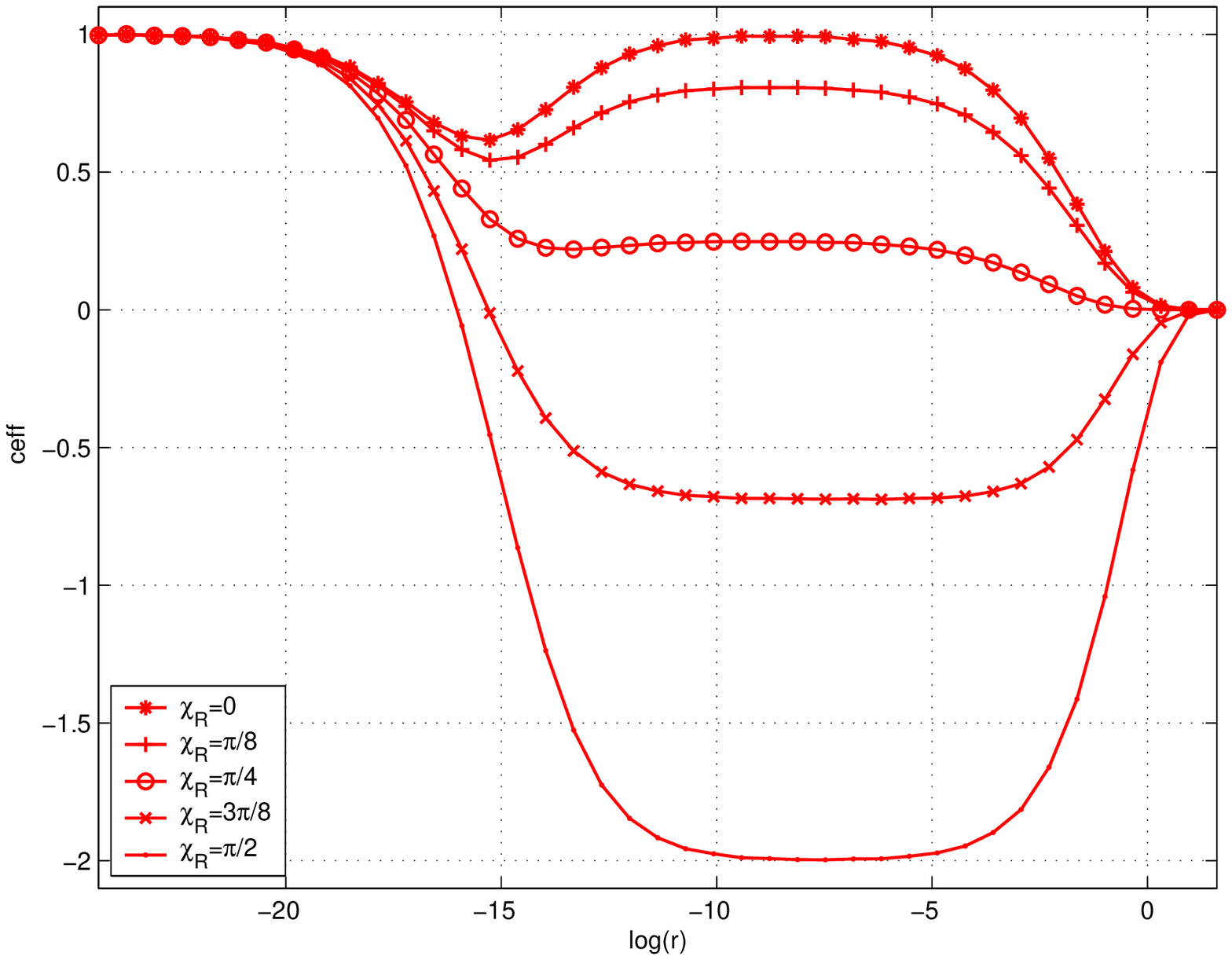}
\\
\parbox{.7\linewidth}{\small 
Figure \ref{fig:ceff_logr_m1000}:
$c_{\rm eff}$~v.s.~$\log(r)$:
$m_B^{(L)} = m_B^{(R)} = 1000$ and 
$\chi_L=\alpha=0$.}
\end{array}
\]
\vskip 8pt
At IR limit $c_{\rm eff} \to 0$. 
At UV limit  $c_{\rm eff} \to 1$  when $\alpha=0$.
However, at the intermediate scale, 
the plot $c_{\rm eff}$~v.s.~$r$ 
(or $c_{\rm eff}$~v.s.~$\log(r)$) 
shows a new feature. 
In figure \ref{fig:ceff_logr_m1000},
$c_{\rm eff}$~v.s.~$\log(r)$ is plotted 
for  $m_B = 1000$ and $\alpha=0$.
(Here $\chi_R=0, \pi/8$ belong to Regime II,
and $\chi_R= 3\pi/8, \pi/2$ to Regime I.)
There appears a plateau around $\log(r) \approx -8$
(here $\vartheta \approx 15 \gg 1$),
which corresponds to the massless limit :
$c_{\rm eff}$~v.s.$\chi$ at this plateau 
reproduces the value in the massless limit
figure \ref{fig:ceff_chi}.
The plateau appears 
around $-\log(r/2) \approx \vartheta/2 \gg 1$,
which is between  $-\log(r /2) \approx \vartheta$
and  $-\log(r /2) \approx 1$.
The bigger $m_B$, the wider the plateau becomes.
This is why the massless limit is easily obtained 
when $m_B \gg 1$. 
The Dirichlet limit of the massless theory 
corresponds to the extreme limit $m_B \to \infty$.

Note that the  massless limit does not coincide 
with the UV limit. 
For a sufficiently large $m_B \gg 1$,
$\vartheta$ provides a large characteristic energy scale.
One may rescale the rapidity around $\vartheta$
to see the massless limit,
which is considered in  Sec.~\ref{sec:Massless}.
UV limit appears when the energy scale is greater 
than $\vartheta$ (length scale is smaller 
than  $-\log(r/2) \approx \vartheta$).

In regime I ($\chi_c < \chi < \pi/2$),
the boundary bound state exists.
Thus $Z_\infty$ has the singularity  
in the physical strip at $\pm i(\pi - \eta_{L/R})$.
In figure~\ref{fig:Zi_contour_theta},
$Z_\infty S$ is shown 
when $\alpha=0$, $m_B=1000 $,
$\chi_L=0$, $\chi_R=2\pi/5$ and $\log(r)\approx-10$.
(Here $\eta_R=4\pi/5$ and $\eta_L=0$).
\vskip 8pt
\[
\begin{array}{c}
\refstepcounter{figure}
\label{fig:Zi_contour_theta}
\epsfxsize=.45\linewidth
\epsfbox{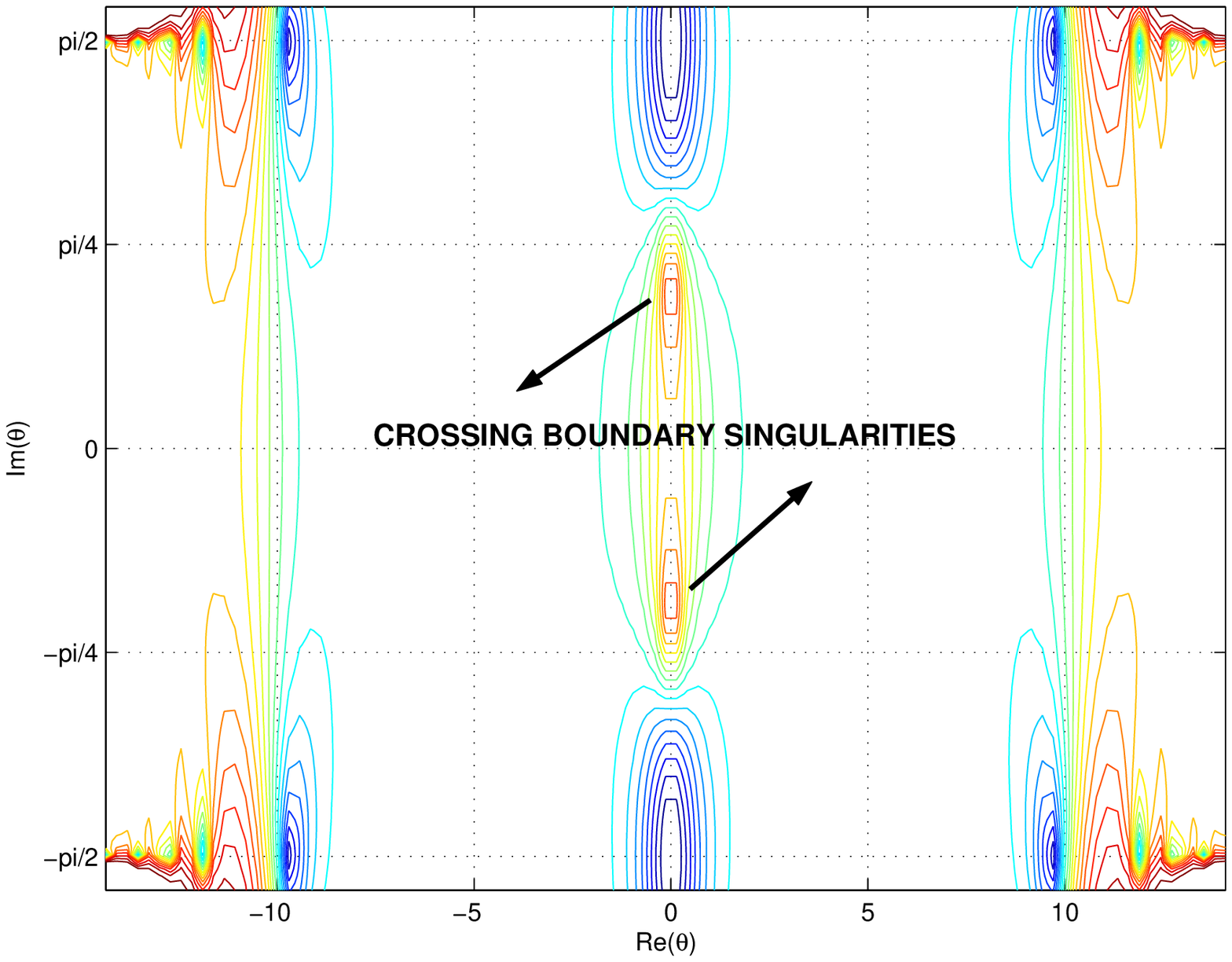}
\qquad
\refstepcounter{figure}
\label{fig:Z_contour_theta}
\epsfxsize=.45\linewidth
\epsfbox{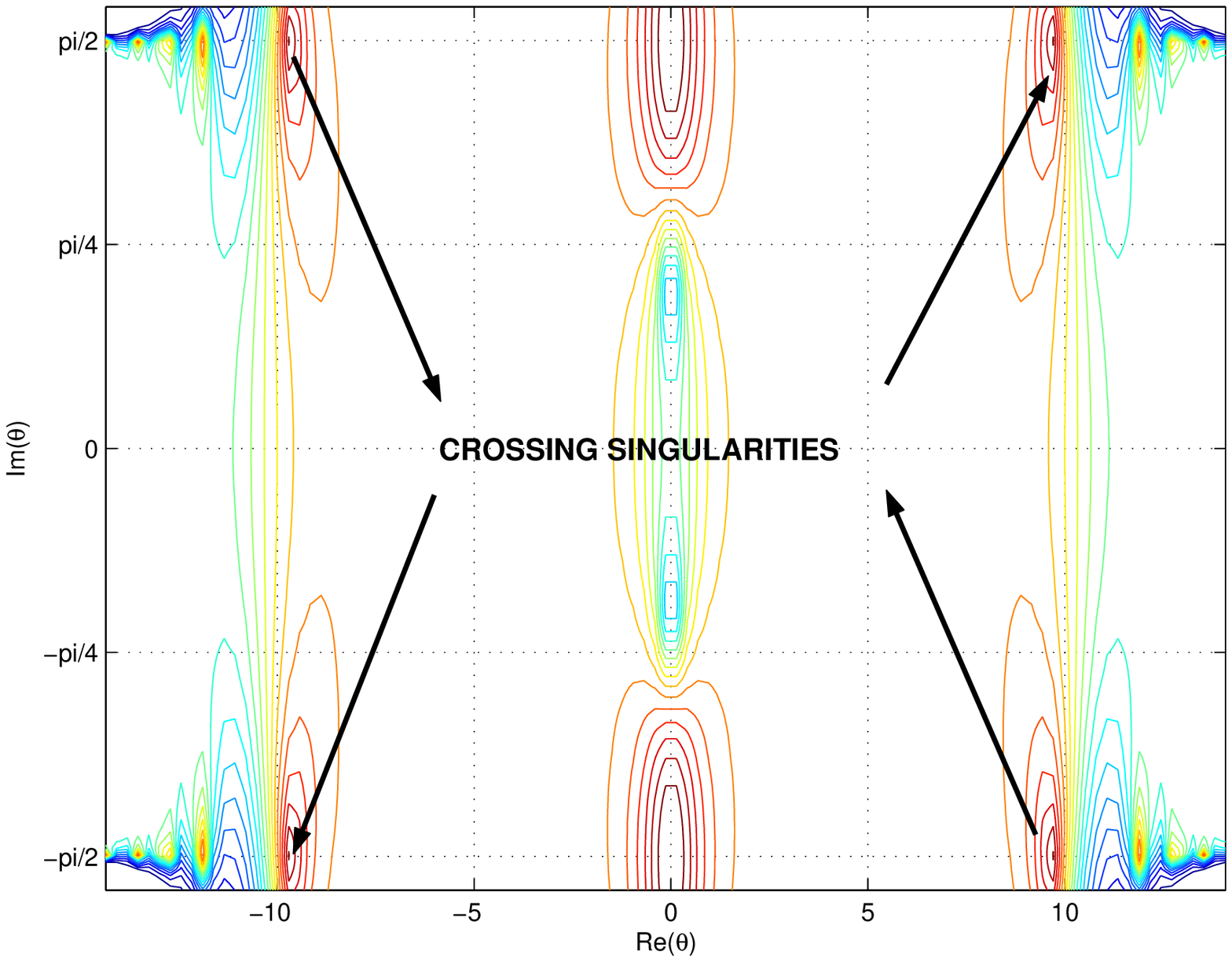}
\\
\qquad
\parbox{.45\linewidth}{\small \raggedright
Figure \ref{fig:Zi_contour_theta}:%
~Contour plot of $Z_\infty S$.
$Z_\infty CS$ is identified with arrows. }
\qquad
\parbox{.45\linewidth}{\small \raggedright
Figure \ref{fig:Z_contour_theta}:%
~Contour plot of $Z_0 S$.
$Z_0 CS$ is identified with arrows.}
\end{array}
\]
\vskip 8pt

$Z_0S$ lies  at  Im($\theta) =\pm \pi/2$
as $r\ll 1 $.
The singularities 
accummulate around at $\theta =\pm (i \pi/2\pm \log(r/2) ) $
along with the isolated one at $\theta =  \pm i \pi/2$
as shown in figure~\ref{fig:Z_contour_theta}.
However, as $r\gg 1$, 
a pair of (crossing) singularities
appear at the pure imaginary rapidity axis
$\pm i \theta_N$ with 
$\theta_N$ real, 
$(\pi-\eta) <\theta_N <\pi/2$.

Regime III ($\pi/2< \chi< (\pi -\chi_c)$)
is the analytically continuation of Regime I
and $c_{\rm eff}$ is modified according to 
the crossing singularities.
Suppose $\chi_L=0$ and 
$\pi/2<\chi_R<(\pi-\chi_c) $.
We need to consider  two cases seperately,
$r$ large and $r$ small. 
When $Z_0CS$ is given as $\pm(\theta_D \pm i \pi/2) $
(the case when $r$ is small),
\beq
c_{\rm eff}  
= -\frac {24r}{\pi} \,
\Big(\cosh {\theta_D} - \sin \eta_R \Big)
+\frac{6r}{\pi^2} -\!\!\!\!\!\!\int _{-\infty}^\infty d \theta
\,e^\theta \, 
\log Z(\theta) \,.
\label{ceff_DD_ext}
\eeq
When $Z_0$ crossing singularity is given as
$\pm i \theta_N$ (the case when $r$ is large),
\beq
c_{\rm eff} (\chi) 
= -\frac {24r}{\pi} \,
\Big(\sin {\theta_N} - \sin \eta_R \Big)
+\frac{6r}{\pi^2} -\!\!\!\!\!\!\int _{-\infty}^\infty d \theta
\,e^\theta \, \log Z(\theta) \,.
\label{ceff_DN_ext}
\eeq
In UV limit, (\ref{ceff_DD_ext}) gives $c_{\rm eff} \to 1$.
In IR limit, as $\theta_N \to \pi- \eta_R $ 
(\ref{ceff_DN_ext}),   $c_{\rm eff} \to 0$.
This behavior is seen in figure \ref{fig:ceff_ext}
when $\chi_R=3\pi/5$.
\vskip 8pt
\[
\begin{array}{c}
\refstepcounter{figure}
\label{fig:ceff_ext}
\epsfxsize=.47\linewidth
\epsfbox{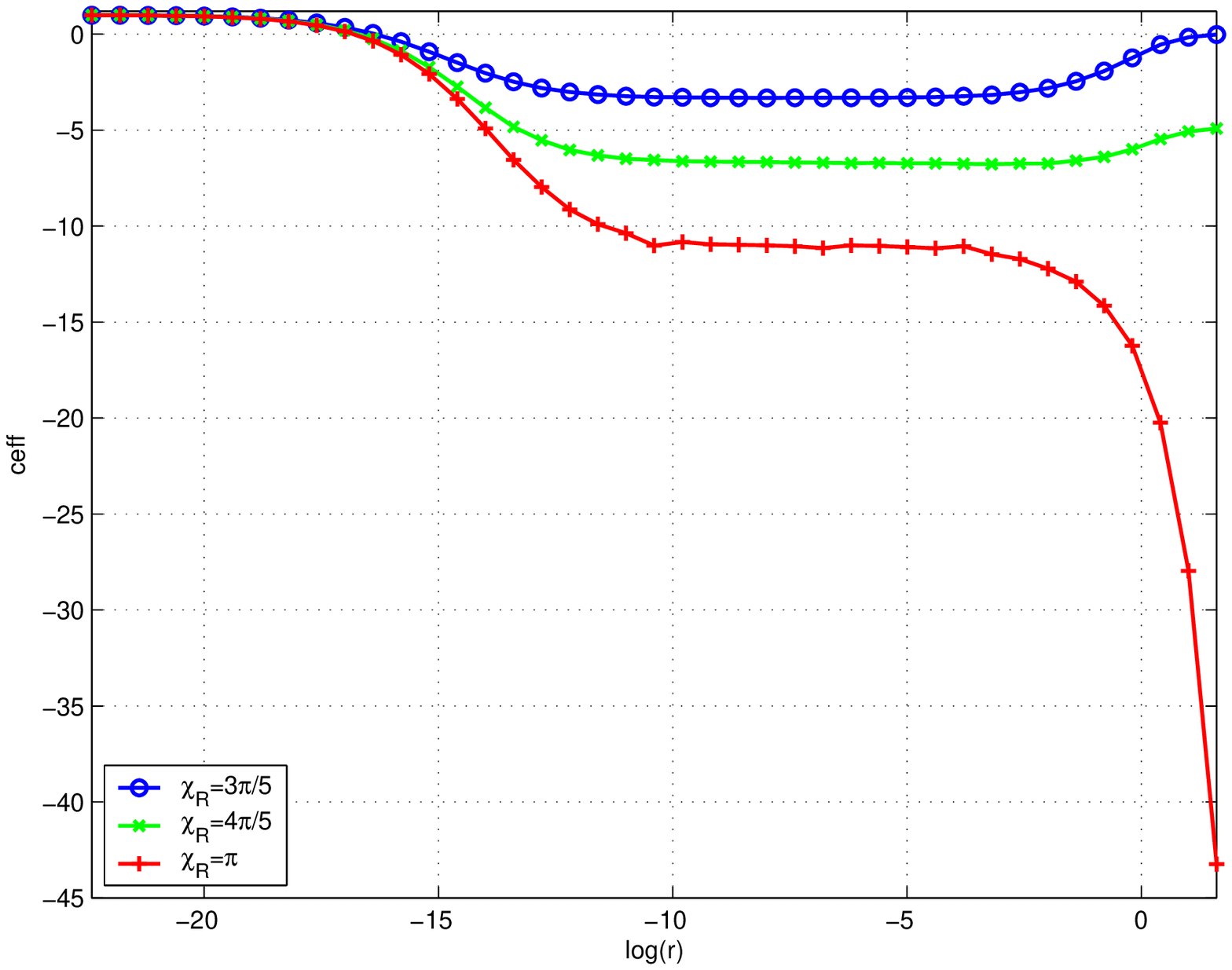}
\quad
\epsfxsize=.47\linewidth
\epsfbox{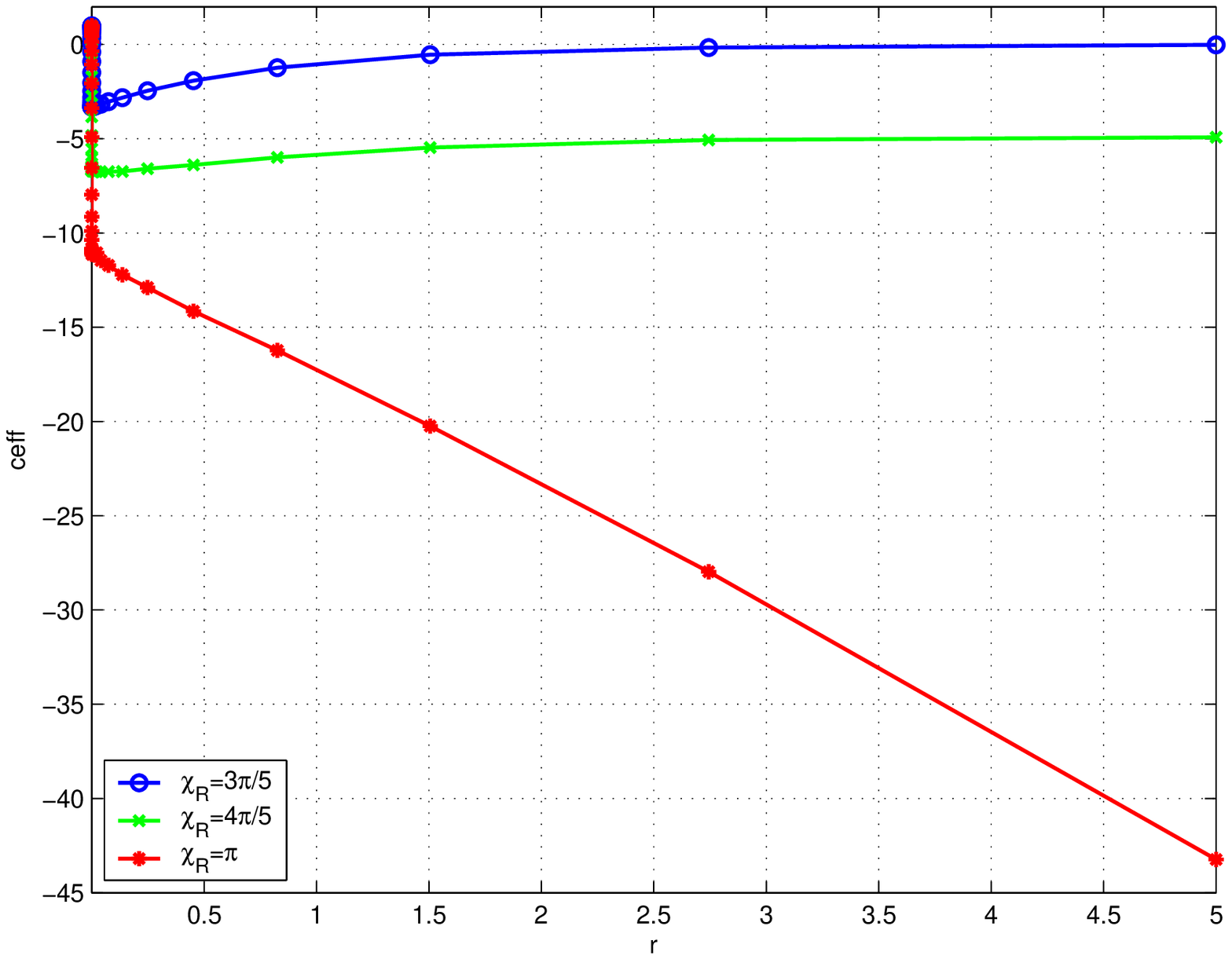}
\\
\parbox{.99\linewidth}{\small \raggedright
Figure \ref{fig:ceff_ext}:%
~$c_{\rm eff}$ v.s. $\log r$ (left) 
and $c_{\rm eff}$ v.s. $ r$ (right)
with $m_B^{(L)}=m_B^{(R)}=1000$.
$\chi_R=3\pi/5$ is in Regime III
and $\chi_R= 4\pi/5, \pi$ are in Regime IV.}
\end{array}
\]
\vskip 8pt
In Regime II, ($ 0 < \chi< (\pi-\chi_c)$),
there is no  boundary bound states and 
therefore no $Z_\infty S$ on the physical strip.
$Z_0S$ is given as Im$(\theta)=\pi/2$.
Analytically continuing to 
Regime IV ($(\pi-\chi_c)< \chi< \pi$),
one has $Z_0CS$  at 
$\pm (\theta_D \pm i\pi/2)$ with $\theta_D $ 
real and positive.
It is noted however,  
even though there is no  boundary bound states,
$Z_\infty CS$ is at $\pm i (\pi -\eta_R)$,
which is outside of physical strip.
Thus, $c_{\rm eff}$ is given
as the same form of (\ref{ceff_DD_ext}):
\beq
c_{\rm eff}  
= -\frac {24r}{\pi} \,
\Big(\cosh {\theta_D} - \sin \eta_R \Big)
+\frac{6r}{\pi^2} -\!\!\!\!\!\!\int _{-\infty}^\infty d \theta
\,e^\theta \, \log Z(\theta) \,.
\eeq

The $r$-dependence is given 
in figure~\ref{fig:ceff_ext}
for $\chi_R= 4\pi/5$ and $\pi$.
In the intermediated scale,
the plateau reproduces the massless limit
in figure \ref{fig:ceff_chi}.
As $r \gg1$, one sees that  $c_{\rm eff}$
has the tendency of decreasing linearly in $r$ 
In fact, $\theta_D \to 0$ at IR limit and 
as  a result
\beq
c_{\rm eff} (\chi) 
= -\frac {24r}{\pi} \, (1 - \sin \eta_R )
+O\left( e^{-2r} \right)\,.
\eeq
The linearity of $c_{\rm eff}$ in $r$ indicates that 
if one changes the sign of one edge,
the free energy density $f(r)$ in IR limit 
becomes bigger than the case with the same sign:
\beq
\Delta f = - \frac{\pi}{24 R} \Delta c_{\rm eff}
= M- M\sin{\eta_R}  \,.
\eeq
The additional free energy 
corresponds to the one soliton mass $M$ 
compensated by the right edge boundary excitation energy 
$M\sin{\eta_R}$.
Thus, one may subtract this IR amount from $c_{\rm eff}$ 
and equivalently, re-normalize the free energy density.

Finally, $\alpha$ effect is vanishingly small at IR limit. 
However, the $\alpha$ effect is enhanced 
at $r$ smaller than the plateau region.
Especially, at UV limit $c_{\rm eff}$ 
is not 1  but depends on values of $\alpha$. 
$c_{\rm eff}$~v.s.~$\alpha$ at UV limit 
reproduces the CFT limit,
which is plotted in figure \ref{fig:ceff_alpha_m100}.
\vskip 8pt
\[
\begin{array}{c}
\refstepcounter{figure}
\label{fig:ceff_alpha_m100}
\epsfxsize=.45\linewidth
\epsfbox{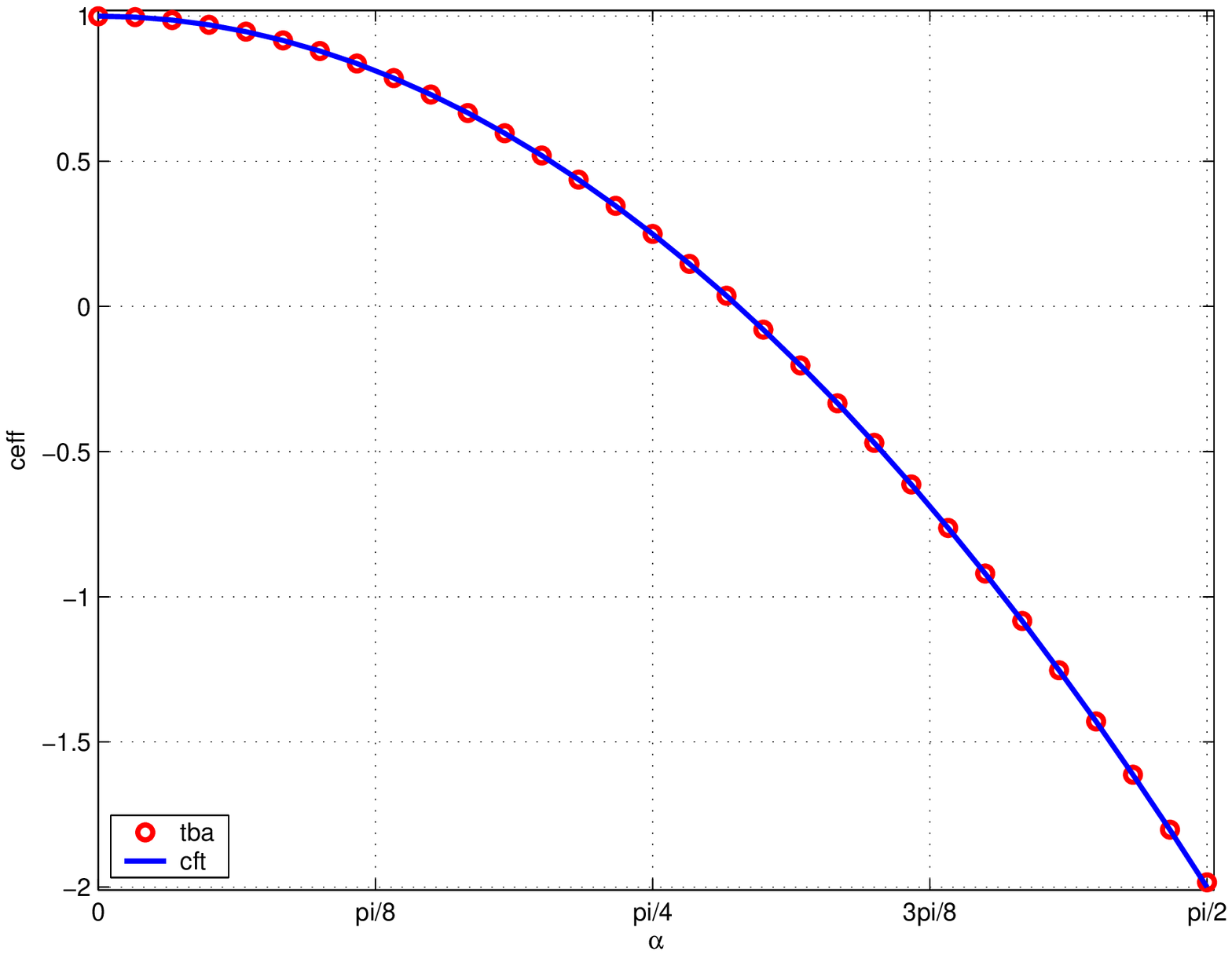}
\\
\parbox{.55\linewidth}{\small 
Figure \ref{fig:ceff_alpha_m100}:
~$c_{\rm eff}$~v.s.~$\alpha$:
$r= \exp(-23)$ and $m_B = 100$.}
\end{array}
\]
\vskip 0.5cm

\section{ND-type boundary condition}
\label{sec:ND}
In this section, we will give 
a few comments when 
combination of N-type at one edge
and D-type at the other edge is given.
Suppose the boundary strength at the left edge
is neglectingly small ($m_B^{(L)}\sim 0$)
and that of the right edge is large
($m_B^{(R)} \gg 1/\sqrt2 $).
In this case the singularity crossing effect 
can be neglected.
(Note that one may choose $\chi< \pi/2$
if $m_B^{(L)}=0$ using the action symmetry).
Still, there remains a plateau, correponding to 
the massless limit in addition to the UV limit.

At UV, one sees the enhanced role of the 
compact boson property,
$c_{\rm eff} \to 1-12(\alpha/\pi)^2$, 
independent of $\chi$. 
At the plateau, 
one may confirm that $c_{\rm eff}$ 
is independent of $\chi$ and $\alpha$ 
and approaches to $-1/2$.
This results in $\Delta= 1/16$,
the dimension of the ground state.
One may see this behavior 
in figure~\ref{fig:ceff_logr_ND_alpha}.
\vskip 8pt
\[
\begin{array}{c}
\refstepcounter{figure}
\label{fig:ceff_logr_ND_alpha}
\epsfxsize=.45\linewidth
\epsfbox{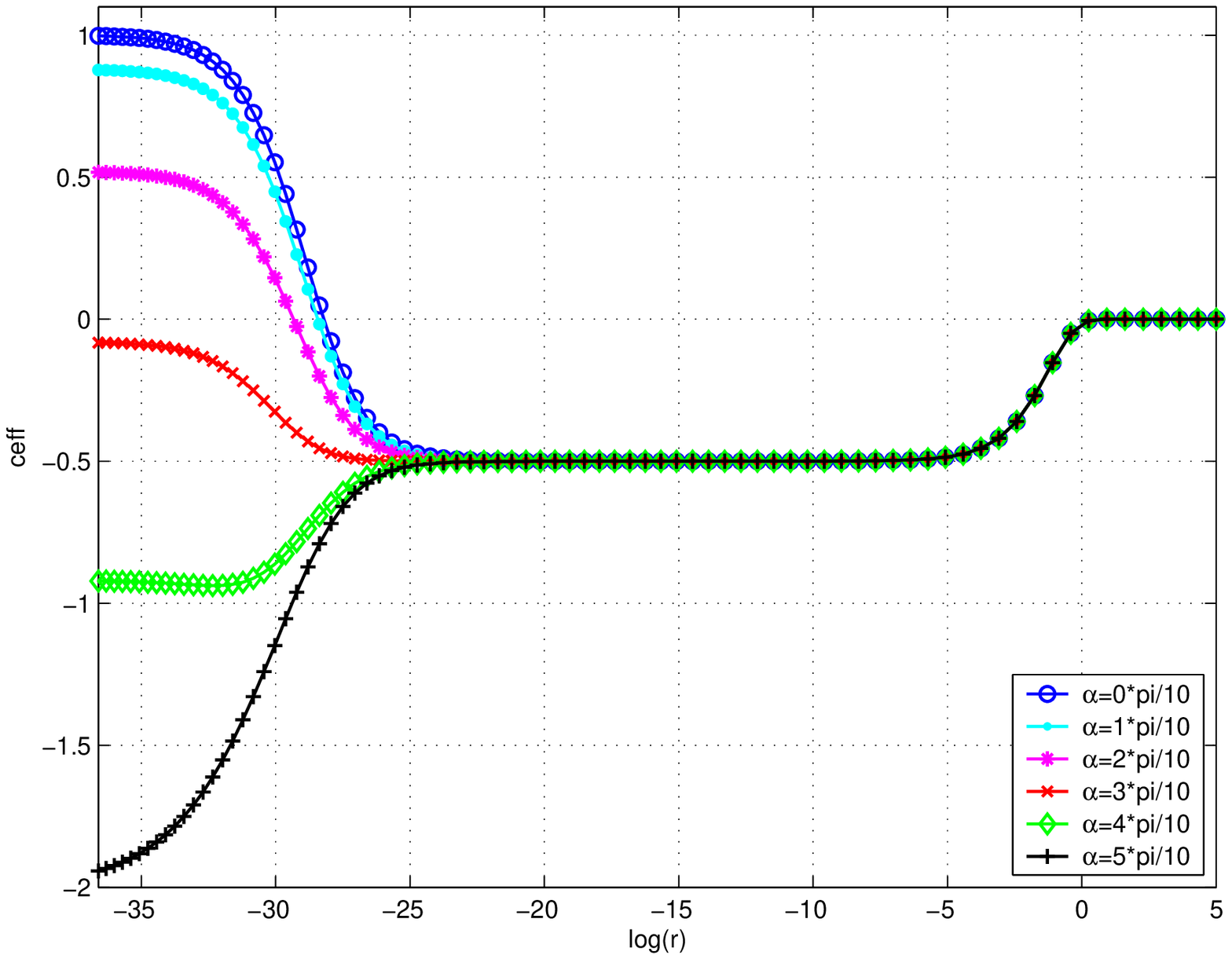}
\qquad
\refstepcounter{figure}
\label{fig:ceff_logr_ND_chi}
\epsfxsize=.45\linewidth
\epsfbox{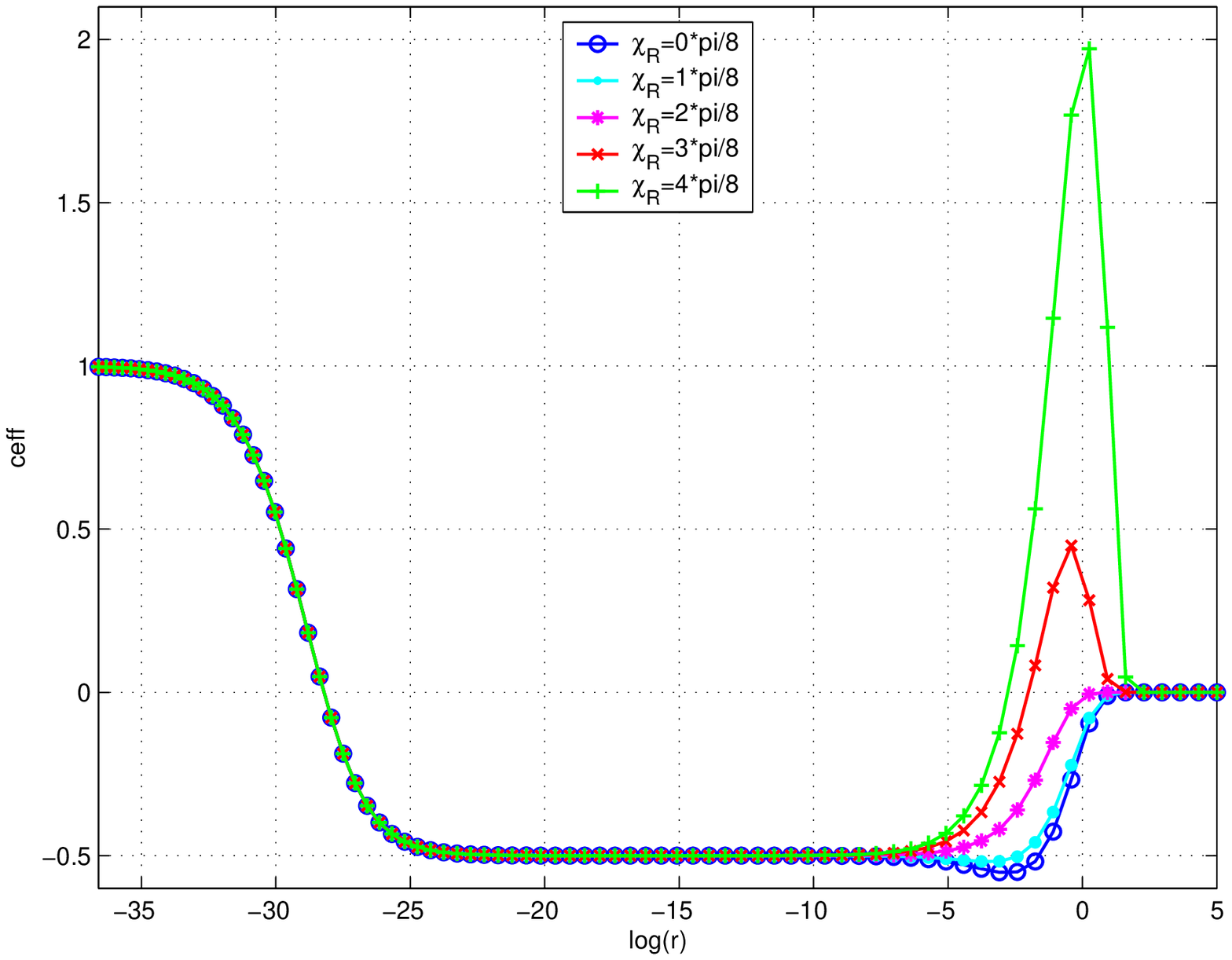}
\\
\parbox{.45\linewidth}{\small 
Figure \ref{fig:ceff_logr_ND_alpha}:
~$c_{\rm eff}$~v.s.~$\log(r)$
with various $\alpha$:
$m_B^{(L/R)}=(10^{-2}, 10^6)$
and $\chi_{L/R}= (0,\pi/4)$ .}
\qquad
\parbox{.45\linewidth}{\small 
Figure \ref{fig:ceff_logr_ND_chi}:
~$c_{\rm eff}$~v.s.~$\log(r)$
with various $\chi$:
$m_B^{(L/R)}=(10^{-2}, 10^6)$
and $\alpha=0$ .}
\end{array}
\]
\vskip 0.5cm
At IR $c_{\rm eff} \to 0$.
In addition, as $r \gg1$, 
there apppears a resonance bump.
The resonance behavior is more enhanced 
as $\chi_R \to \pi/2$,
since the singularities move close to the 
real rapidity axis.
This is clearly seen in figure~\ref{fig:ceff_logr_ND_chi}.

If the boundary strength $m_B$ is not 
in the extreme case, 
the singularity crossing 
contribution is not significant. 
When one edge is in $0<\chi<\pi/2$ 
and the other edge in $\pi/2 <\chi<\pi$,
using the action symmetry 
one may re-arrange the N-type boundary in 
$\pi/2 <\chi<\pi$ (Regime III)
and D-type boundary in 
$0<\chi<\pi/2$ (Regime II)
so that one counts in the singularity 
crossing from the  N-type boundary,
which will give the ground state contribution.
 
\vskip 0.5cm 

\section{Conclusion}
\label{sec:Conclusion}

To summarize, we consider the free fermion limit of the 
massive boundary sine Gorodon model on a strip 
and investigate the boundary parameter effect
on the RG flow of $c_{\rm eff}$.
For a certain range of parameters,
the flow shows 
$c_{\rm eff} = 0$ at IR limit,
and  $c_{\rm eff} \to 1 - 12 (\alpha /\pi)^2 $
at UV limit. 
However, depending on the paramter ranges
the RG flow is not monotonic and 
and shows much more diverse effects 
at the intermediate scale.
All the anomalous behavior 
originates from
the singularity structure on the 
complex rapidity space. 

In NN-type boundary condition 
($m_B^{(L/R)} < 1\sqrt2$),
UV limit and massless limit coincide.
In the intermediate scale a resonance
appears due to the singularity close to 
the real rapidity axis.
If one edge is in $0< \chi <\pi/2$
(Regime I), 
and the other edge is in $\pi/2 < \chi <\pi$
(Regime III),
then the singularity crossing is to be counted
to modify the $c_{\rm eff}$ formula. 

In DD-type boundary condition 
($m_B^{(L/R)} > 1\sqrt2$),
UV limit does not coincide with the massless limit. 
The massless limit appears 
as the plateau in $c_{\rm eff}$~v.s.~$\log(r)$,
which is characterized 
by the scale $r \sim e^{-\vartheta}$.
The Dirichlet limit of this plateau
reproduces the massless result,
$c_{\rm eff} \to 1 - 12 (\chi /\pi)^2 $
with $\chi=\chi_R-\chi_R$.

If one edge is in $0< \chi <\pi/2$
(Regime I or II)
and the other edge is in $\pi/2 < \chi <\pi$
(Regime III or IV),
then the singularity crossing modifies
the $c_{\rm eff}$ formula. 
Especially when one edge is in Regime II
($0< \chi <\pi/2$ and $0<\eta < \pi/2 $), 
and the other edge in Regime IV
($\pi/2 < \chi <\pi $ and $0<\eta < \pi/2 $), 
then $c_{\rm eff}$ is linearly decreasing in the 
scale $r$,
which indicates the opposite sign of boundary terms
excites the one-particle state of the same sign of 
boundary terms.  
Similar behavior is obtained for Ising model 
when opposite boundary magnetic fields 
are applied to both edges. 
The magnetic field should not be too strong
($-1<k<0$), otherwise the opposite fields spoil the 
field theory and result in the complex free energy.

In ND-type, 
($m_B^{(L)} < 1\sqrt2$ and $m_B^{(R)} > 1\sqrt2$),
still UV limit and massless limit have their own 
domain. In the extreme limit 
($m_B^{(L)} \to 0 $ and $m_B^{(R)} \gg 1$),
one may see that UV limit is the compact bosonic 
CFT with  $c_{\rm eff} \to 1-12(\alpha/\pi)^2$.
On the other hand, the massless limit gives 
$c_{\rm eff} \to -1/2$. 

The similar singularity effect on RG flow is expected to be
persistent when bsG goes beyond free Fermi limit.
However, the presence of more species of particles 
and boundary bound states makes RG flow more complicated
and the numerical analysis will be more 
difficult due to the nemerical instability.

It is noted that the singularity analysis 
for the case of opposite boundary terms
will be useful to study other boundary field theories. 
In Ising model, the magnitude of opposite boundary fields 
is limited since the field theory breaks down. 
This does not happen in bsG. 
On the other hand, in boundary affine Toda field theory, 
the boundary actions are given in 3 kinds only.
Among them the (-)boundary action is not well understood
\cite{TODA1,TODA2,TODA3}. 
It seems that the complete understanding 
of the boundary action requires the proper study of 
the singularity structure 
at the presence of (-)boundary action condition.

\vskip 8pt
\section*{\bf Acknowledgement}

The author thanks Z.~Bajnok, P.~Dorey,
R.~Nepomechie, F. Ravanini 
and R.~Tateo for valuable discussions,
T.~J.~Lee for the work at the initial stage,
and APCTP topical program during which this work initiated. 
This work is supported in part 
by Korea Research Foundation 2002-070-C00025.   

\section*{Appendix}

Ising model's reflection amplitude is given as
\beq
R_k(\theta) = i \tanh\Big(\frac{i\pi}4 - \frac{\theta}2 \Big)\,
\frac{k- i\sinh \theta}{k- i\sinh \theta}
\eeq
where $k= 1 - h^2$
with $h$  the boundary magnetic field.
(Note that $h^2$ is scaled by $2M$ 
where $M$ the particle mass 
so that $h$ is a dimensionless number.)
Thus 
\beq
K(\theta) =  i \tanh\Big(\frac{\theta}2 \Big)\,
\frac{k+\cosh \theta}{k- \cosh \theta}
\eeq
and  the fugacity becomes
\beq
\lambda = \overline{K_L} K_R
= \tanh^2 \Big(\frac{\theta}2 \Big)\,
\frac{k_L+\cosh \theta}{k_L- \cosh \theta}
\frac{k_R+\cosh \theta}{k_R- \cosh \theta}\,.
\eeq
The effective central charge is given as 
\cite{ising-TBA}
\beq
c_{\rm eff}  
=  \frac{6r} {\pi^2 } \int_{-\infty}^\infty d \theta \,
\cosh \theta \,  \log Z_{\rm Ising}(\theta)
\equiv  c-24\Delta  \,,
\eeq 
where 
$\epsilon =2r\cosh \theta$ with $r=RM$.
$Z_{\rm Ising}(\theta)
= \left ( 1 + \lambda(\theta)\, 
e^{- \epsilon(\theta)} \right)$
satisfies the periodicity,
$Z_{\rm Ising}(\theta + i 2 \pi) 
= Z_{\rm Ising}(\theta)$
and 
$Z_{\rm Ising}(\theta + i  \pi) 
= Z_{\rm Ising}(\theta) \,
e^{\epsilon_0 (\theta)} / \lambda_0(\theta)$.

According to singularity structure,
the domain of parameters are classified into 
three regimes: $0<k<1$, $-1<k<1$ and $k<-1$.
In each regime, the magnetic field $h$ and $-h$ is 
to be distinguished.
In fact,  $h=0$ (or $k=1$)
is the singular point in the parameter space $h$ 
since one of $Z_0S$ lies 
at real rapidity axis. The domains $h>0 $ and $h<0$ 
are connected through the branch singularity.
Thus two domains are not equivalent due to 
the existence of crossing singularities.
 
Regime $0<k<1$ is the weak magnetic field regime
($0<h^2<1$)
similar to the N-type in the text.
There is a boundary bound state,
which makes a pair of $Z_\infty CS$ 
at $\pm i \xi$ where 
$0< \xi <\pi/2$ and $\cos \xi =k$. 
A pair of $Z_0 CS$ 
at $\pm i \theta_N $ with $0< \theta_N <\xi$
is  squeezed between 
the pair of $Z_\infty CS$.
$c_{\rm eff}$ is modified 
when the boundary magnetic fields are of opposite sign
($h_L h_R<0$),
\beq
c_{\rm eff} 
= -\frac {24r}{\pi} \,
\Big(\sin \theta_N - \sin \xi \Big)
+\frac{6r}{\pi^2} -\!\!\!\!\!\!\int _{-\infty}^\infty d \theta
\,e^\theta \, 
\log Z_{\rm Ising}(\theta)  \,.
\label{ceff_ising_N_ext}
\eeq
\vskip 8pt
\[
\begin{array}{c}
\refstepcounter{figure}
\label{fig:ceff_logr_ising_N}
\epsfxsize=.5\linewidth
\epsfbox{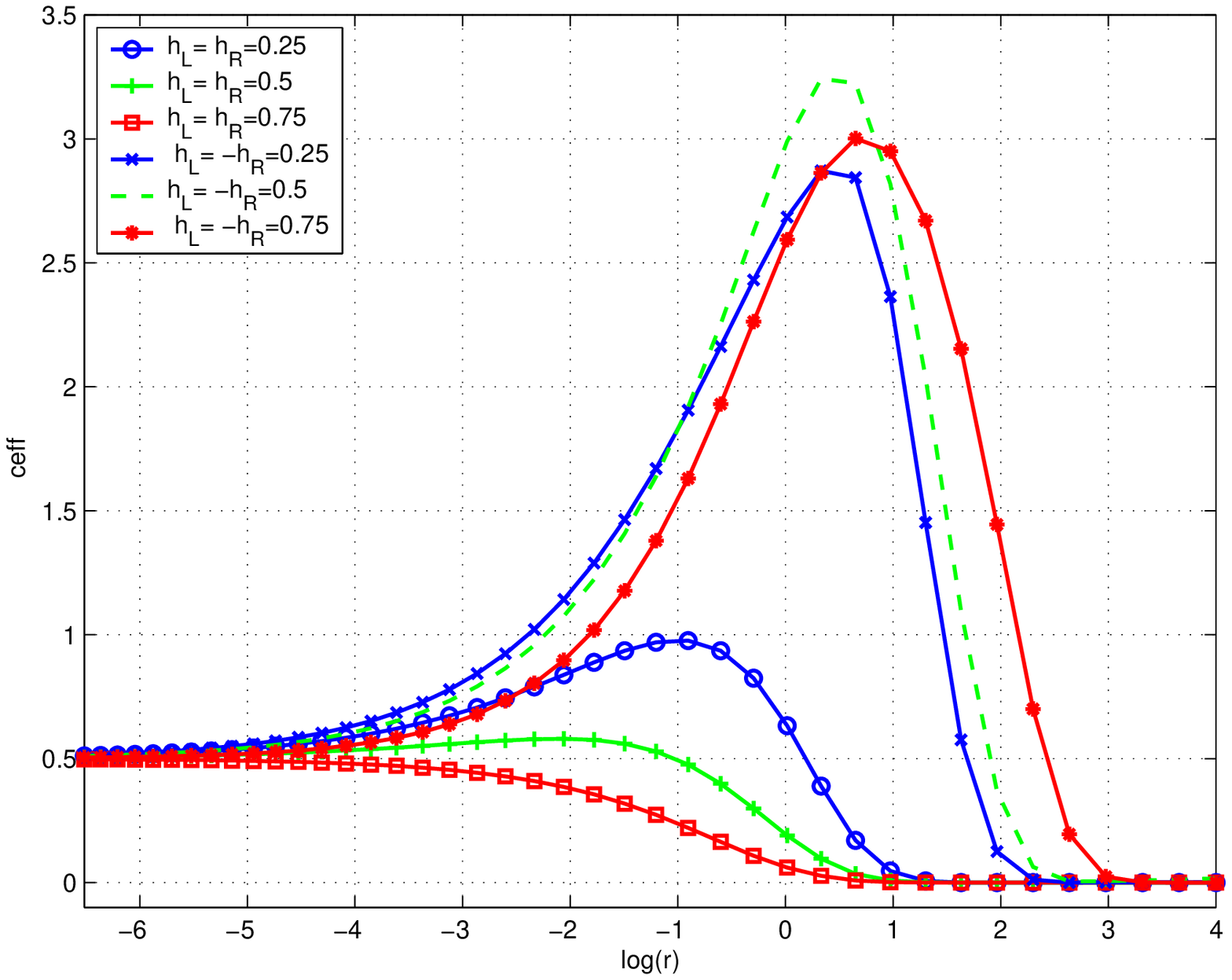}
\\
\qquad
\parbox{.8\linewidth}{\small 
Figure \ref{fig:ceff_logr_ising_N}:%
~$c_{\rm eff}$~v.s.~$\log(r)$ for $0<k<1$
(weak boundary magnetic field).}
\end{array}
\]
\vskip 8pt

In regime $-1<k<0$
(boundary magnetic field $1<h^2<2$), 
there is no boundary bound state.
Nevertheless, $Z_\infty CS$ 
exists at $\pm i (\xi + \pi/2)$  ($0<\xi<\pi/2$),
outside of the physical strip.
All $Z_0 CS$ lie at Im$(\theta)=\pi/2$ and
$Z_0 CS$ is given as $\pm(\theta_D \pm i\pi/2)$.
When the boundary magnetic fields are of opposite sign,
\beq
c_{\rm eff} 
= -\frac {24r}{\pi} \,
\Big(\cosh \theta_D - \cos \xi \Big)
+\frac{6r}{\pi^2} -\!\!\!\!\!\!\int _{-\infty}^\infty d \theta
\,e^\theta \, \log Z_{\rm Ising}(\theta) \,.
\label{ceff_ising_M_ext}
\eeq
In figure \ref{fig:ceff_logr_ising_M}
$c_{\rm eff}$ with $h_Lh_R>0$ 
is plotted.
The case  with $h_Lh_R<0$  is 
contrasted with that of same sign
in figure \ref{fig:ceff_logr_ising_M_whole}
where the curves with the same sign 
are nearly overlapped and are not 
distinguished each other. 
In all cases, there is no resonance. 
\vskip 8pt
\[
\begin{array}{c}
\refstepcounter{figure}
\label{fig:ceff_logr_ising_M}
\epsfxsize=.48\linewidth
\epsfbox{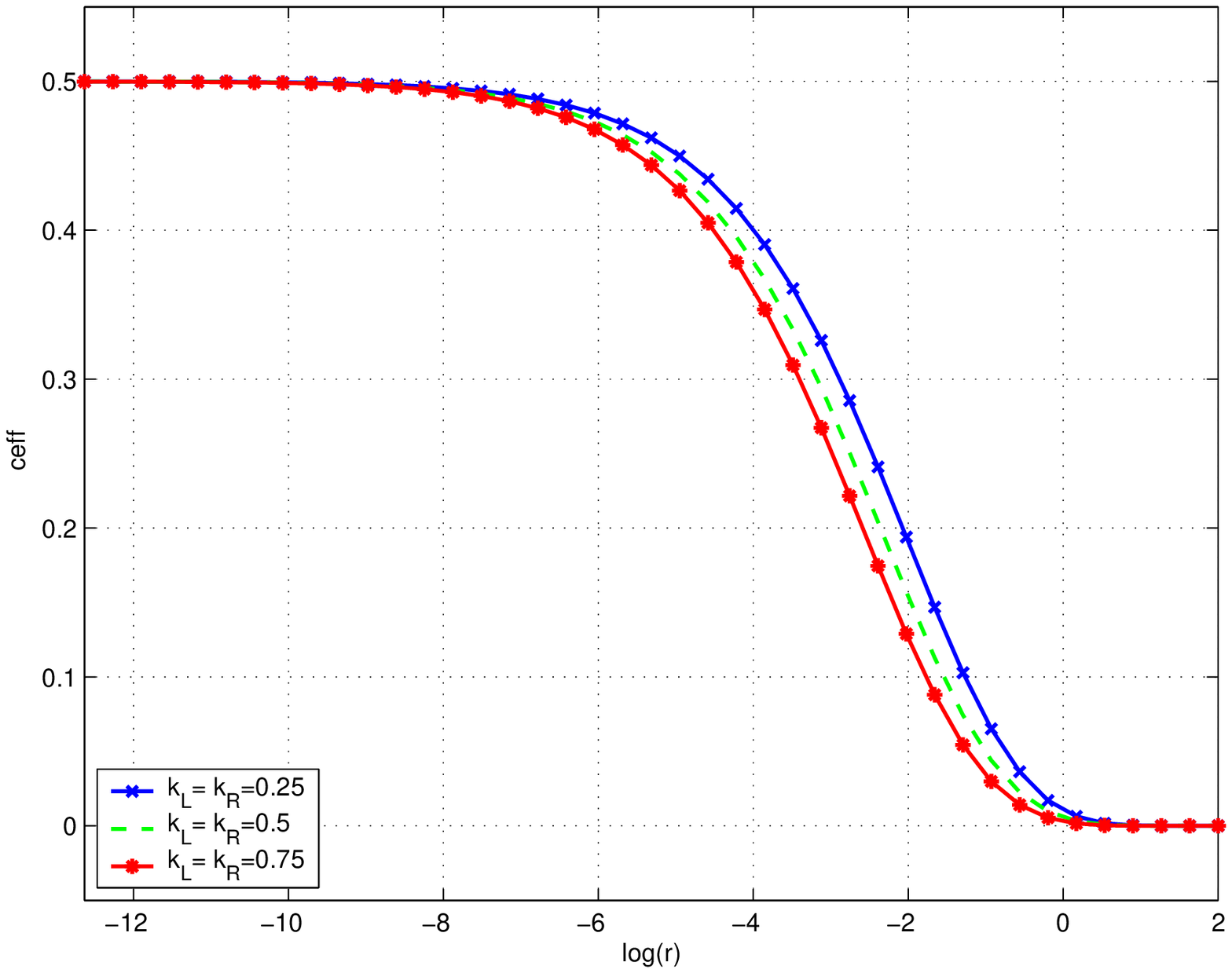}
\qquad
\refstepcounter{figure}
\label{fig:ceff_logr_ising_M_whole}
\epsfxsize=.48\linewidth
\epsfbox{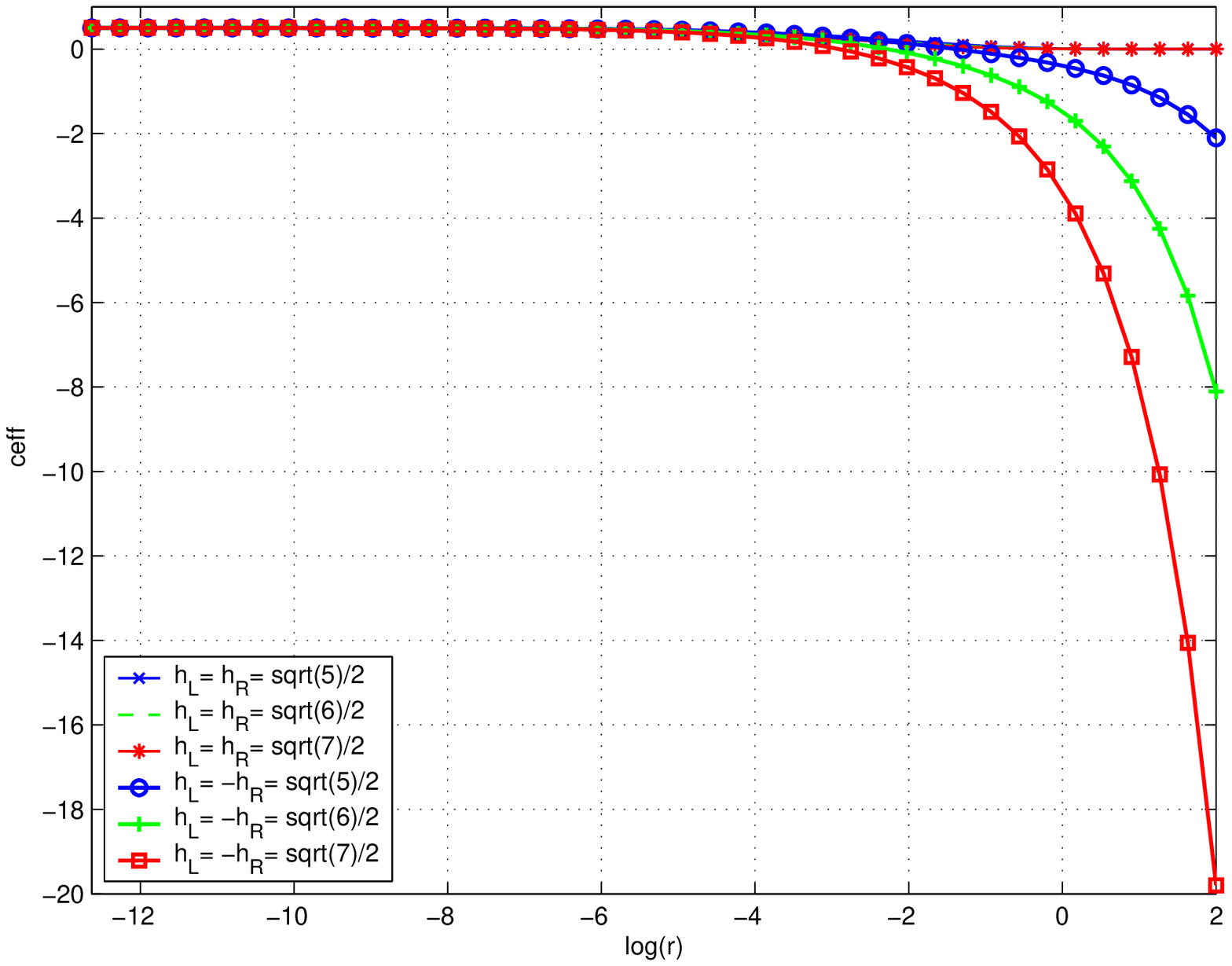}
\\
\qquad
\parbox{.48\linewidth}{\small \raggedright
Figure \ref{fig:ceff_logr_ising_M}:%
~$c_{\rm eff}$~v.s.~$\log(r)$ with $-1<k<0$.
Boundary magnetic field of both edges 
are with same sign.}
\qquad
\parbox{.48\linewidth}{\small \raggedright
Figure \ref{fig:ceff_logr_ising_M_whole}:%
~$c_{\rm eff}$~v.s.~$\log(r)$ with $-1<k<0$.
Boundary magnetic field of both edges 
are with opposite sign}
\end{array}
\]
\vskip 8pt

Note that in figure
\ref{fig:ceff_logr_ising_M_whole},
$c_{\rm eff}$ decreases linearly in $r$ as $r \gg 1$
and when opposite boundary fields are given.  
(At IR limit, $\theta_D=0$ and the integrated part vanishes
in (\ref{ceff_ising_M_ext}) so that the linear 
coefficient is proportional to $(1-\cos \xi)$.)
This trend of linearity is clearly seen
in figure \ref{fig:ceff_r_ising_M_whole}.
\vskip 8pt
\[
\begin{array}{c}
\refstepcounter{figure}
\label{fig:ceff_r_ising_M_whole}
\epsfxsize=.5\linewidth
\epsfbox{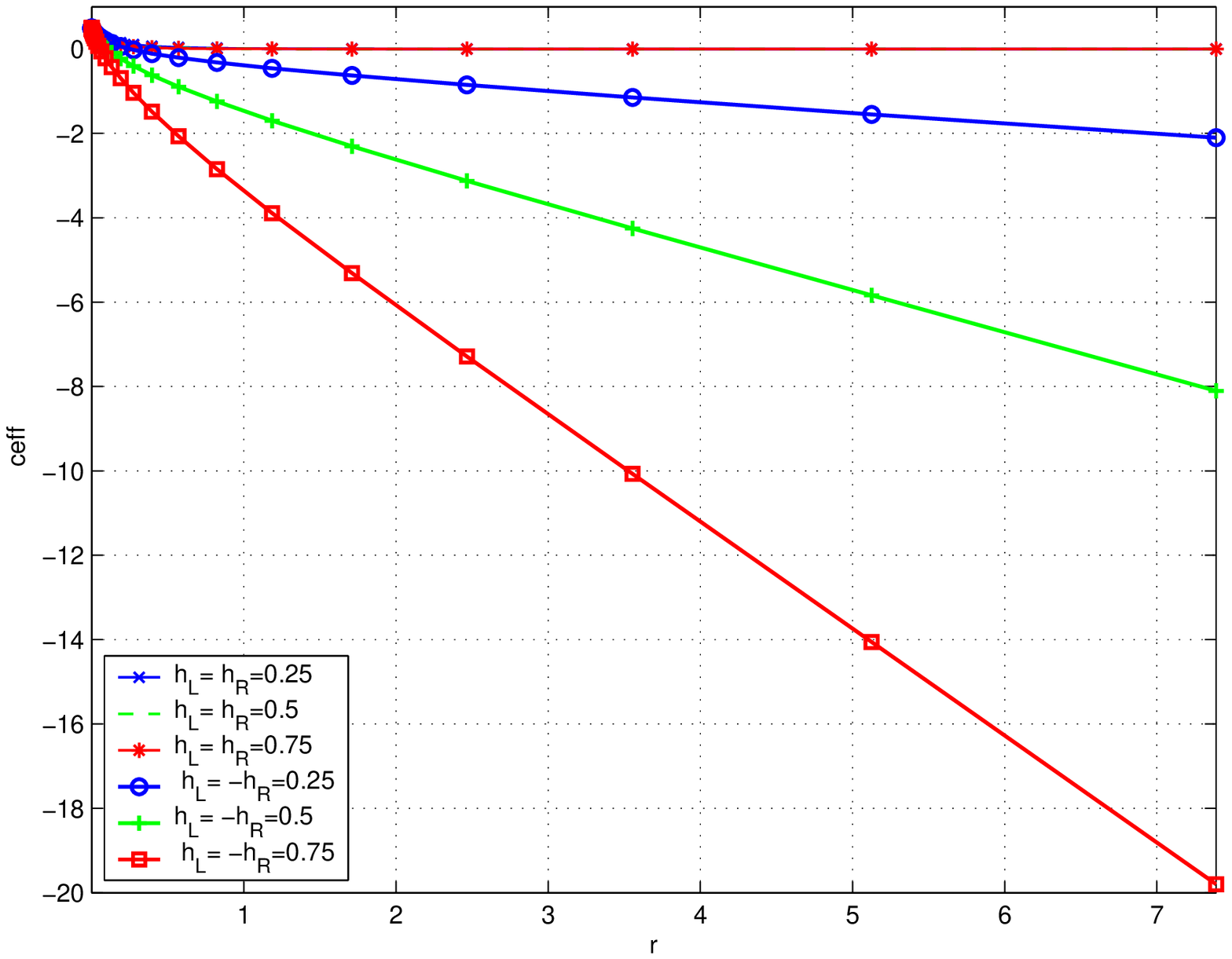}
\\
\qquad
\parbox{.45\linewidth}{\small 
Figure \ref{fig:ceff_r_ising_M_whole}:%
~$c_{\rm eff}$~v.s.~$r$ for $-1<k<0$.}
\end{array}
\]
\vskip 8pt

The linearity of $c_{\rm eff}$ 
in the Regime $-1<k<0$ 
shows the similarity to D-type of bsG.
However, unlike bsG
the plateau does not appear in this Regime
but in Regime $k<-1$ 
(boundary magnetic field $h^2 >2$).
As given in figure \ref{fig:ceff_logr_ising_D},
the stronger the boundary field becomes, 
the wider appears the plateau region 
which is identified as the massless limit.
The massless limit is distinguished from 
the UV limit even though 
$c_{\rm eff} \to 1/2 $ in both limits. 
When $k>-1$, the massless limit coincides with 
the UV limit.
\vskip 8pt
\[
\begin{array}{c}
\refstepcounter{figure}
\label{fig:ceff_logr_ising_D}
\epsfxsize=.6\linewidth
\epsfbox{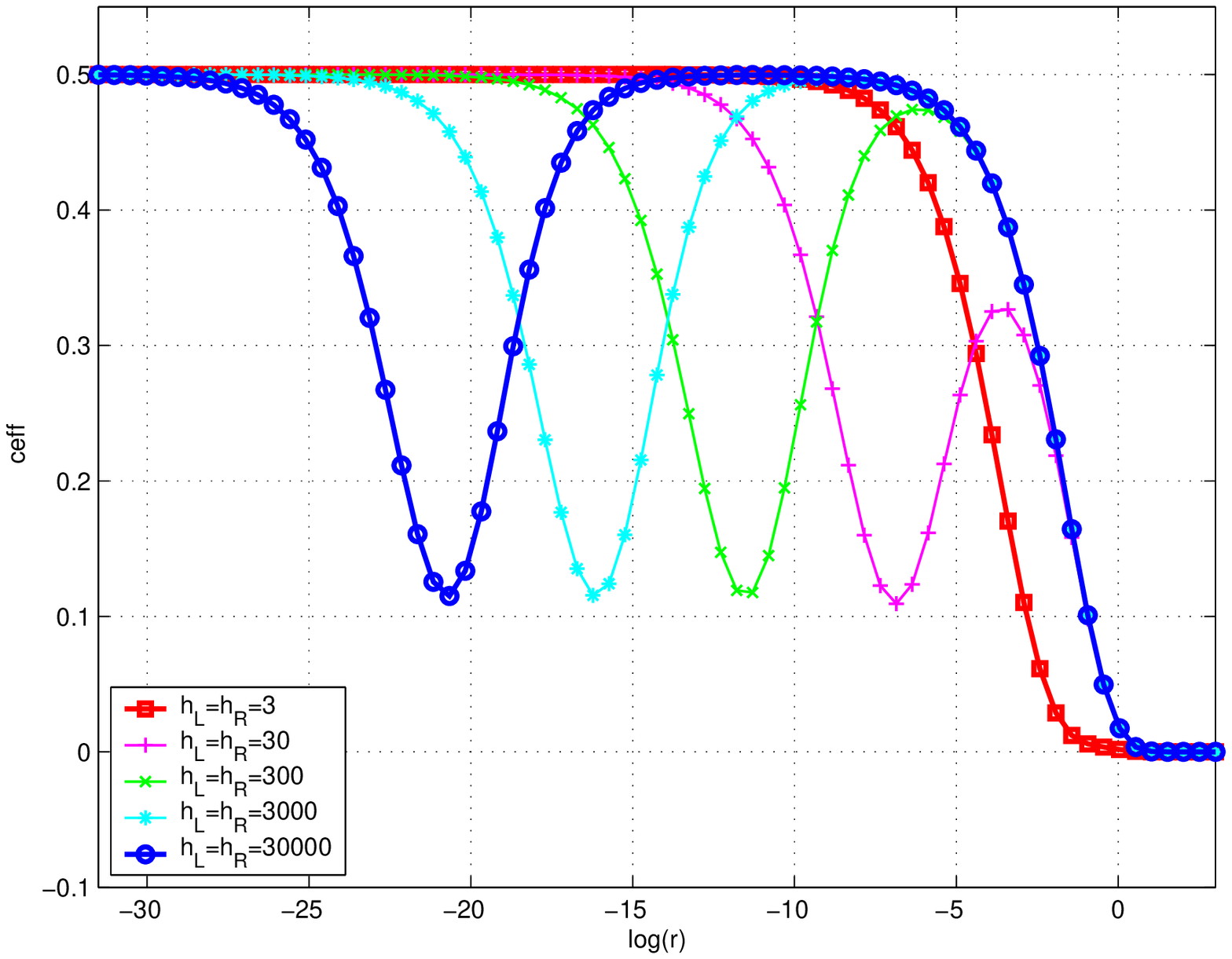}
\\
\qquad
\parbox{.4\linewidth}{\small 
Figure \ref{fig:ceff_logr_ising_D}:%
~$c_{\rm eff}$~v.s.~$\log(r)$ for $k<-1$.}
\end{array}
\]
\vskip 8pt

When $k<-1$, the opposite boundary magnetic field 
makes the field theory unstable.
The breakdown is due to 
$Z_\infty CS$ at $\pm (\xi \pm i\pi)$,
where  $\cosh \xi= |k|$ with real $\xi>0$.
The singularity lies outside of the physical strip
and becomes `resonant poles'.
This crossing singularities result in 
a complex-valued $c_{\rm eff}$
and thus, free energy dendsity becomes complex valued :
\beq
c_{\rm eff} 
= -\frac {24r}{\pi} \,
\Big(\cosh \theta_D + i  \sinh \xi \Big)
+\frac{6r}{\pi^2} -\!\!\!\!\!\!\int _{-\infty}^\infty d \theta
\,e^\theta \, \log Z_{\rm Ising}(\theta) \,,
\label{ceff_ising_D_ext}
\eeq
where $Z_0 CS$ is at $\pm(\theta_D \pm i\pi/2)$.

Finally, if one edge is given with 
a vanishing boundary magnetic field  $k \to 1$
and the other edge with  a strong one $k \to - \infty$,
the plateau appears corresponding to CFT limit
of the mixed boundary condition, whose 
$c_{\rm eff}=-1$
and the ground state has the conformal dimension 
1/16. (See figure \ref{fig:ceff_logr_ising_ND}.)
\vskip 8pt
\[
\begin{array}{c}
\refstepcounter{figure}
\label{fig:ceff_logr_ising_ND}
\epsfxsize=.5\linewidth
\epsfbox{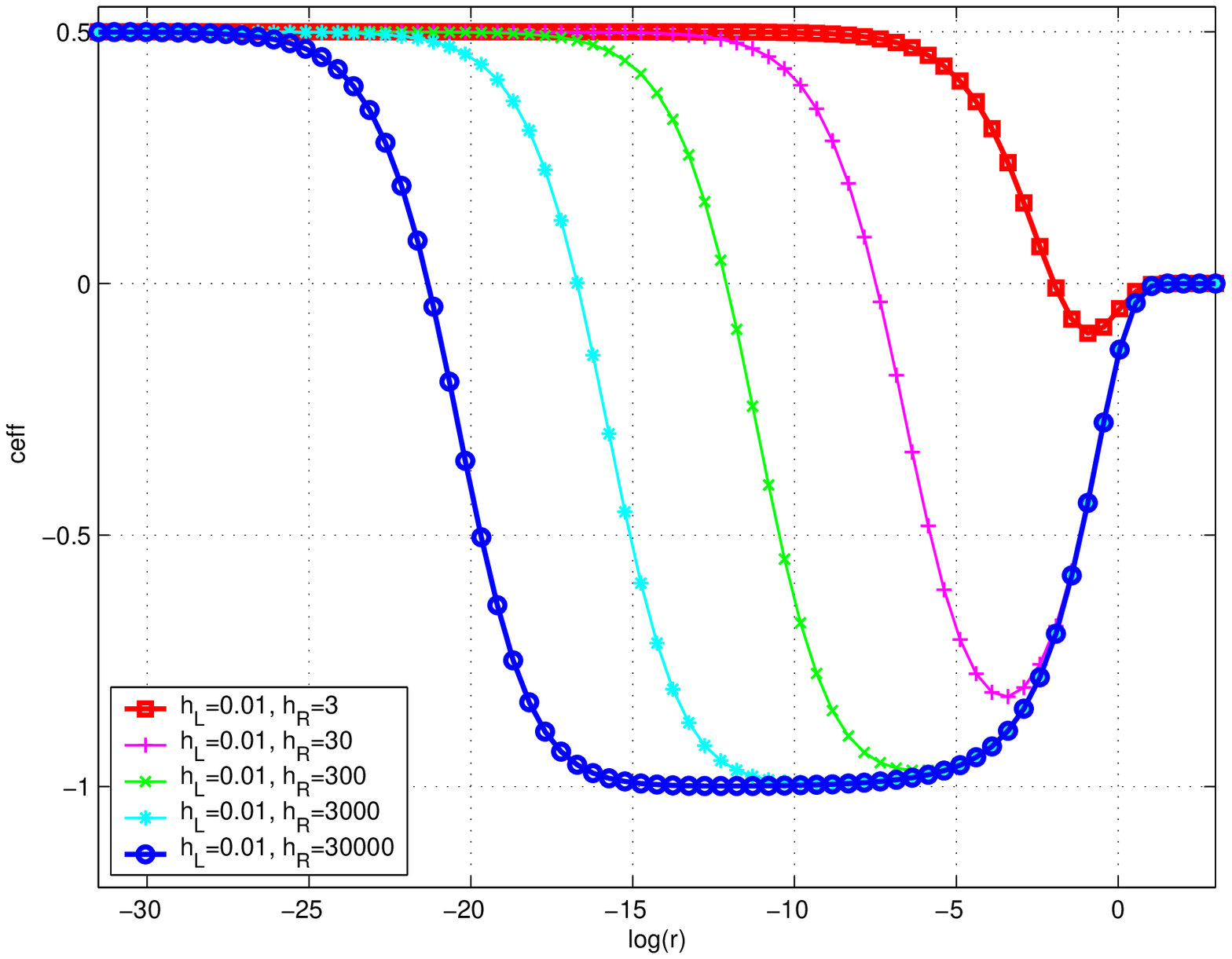}
\\
\qquad
\parbox{.99\linewidth}{\small 
Figure \ref{fig:ceff_logr_ising_ND}:%
~$c_{\rm eff}$~v.s.~$\log(r)$.
A weak (strong) boundary field is given at 
one (the other) edge.}
\end{array}
\]


\begin{thebibliography}{99}
\bibitem{c-theorem} 
A.~B.~Zamolodchikov, 
Int. J. Mod. Phys. {\bf A4} (1989) 4235.
\bibitem{TBA} 
Al. B. Zamolodchikov, 
Nucl. Phys. {\bf B342} (1990) 695.
\bibitem{YY} C.~N.~Yang and C.~P.~Yang, 
J. Math. Phys. {\bf 10} (1969) 1115.
\bibitem{TBA-RG1}
Al. B. Zamolodchikov, 
Nucl. Phys. {\bf B358} (1991) 524.
\bibitem{TBA-RG2}
T.~R.~Klassen and E.~Melzer,
Nucl.\ Phys.\ {\bf B400} (1993) 547.
\bibitem{TBA-RG3}
F.~Ravanini, R.~Tateo, A.~Valleriani,
Int.\ J.\ Mod.\ Phys.\ {\bf A8} (1993) 1707.
\bibitem{TBA-RG4}
P.~A.~Pearce, L.~ Chim, C.~Ahn, 
Nucl.\ Phys.\ {\bf B601} (2001) 539;
{\bf  B660} (2003) 579.
\bibitem{TBA-RG5}
C.~Ahn, C.~Kim, C.~Rim and Al. B. Zamolodchikov, 
Phys. Lett. {\bf B 541} (2002) 94.
\bibitem{bsG-GZ} S.~Ghoshal and A.~B.~Zamolodchikov, 
Int. J. Mod. Phys. {\bf A9} (1994) 3841
(Erratum-ibid {\bf A9} (1994) 4353; 
S.~Ghoshal, Int. J. Mod. Phys. {\bf A9} (1994) 4801.
\bibitem{bLY}
P.~Dorey and R.~Tateo, Nucl. Phys. {\bf B482} (1996) 639;
Nucl. Phys. {\bf B515} (1998) 575.
\bibitem{bsG-CSS1}
J.-S.~Caux, H.~Saleur and F.~Siano, 
Phys. Rev. Lett. {\bf 88} (2002) 106402.
\bibitem{bsG-R}
T.~Lee and C.~Rim,
Nucl. Phys. {\bf B672} (2003) 487.
\bibitem{bsG-CSS2}
J.-S.~Caux, H.~Saleur and F.~Siano, 
Nucl. Phys. {\bf B672} (2003).
\bibitem{R-TBA}
A.~LeClair, G.~Mussardo, H.~Saleur and S.~Skorik,
Nucl. Phys. {\bf B453} (1995) 581. 
\bibitem{bsG-AN} 
C.~Ahn and R.~Nepomechie, 
Nucl.\ Phys.\ {\bf B676} (2004) 637.
\bibitem{bsG-free}
M.~Ameduri, R.~Konik, A.~LeClair,
Phys.\ Lett.\ {\bf B354} (1995) 376.
\bibitem{mandelstam}
S.~Mandelstam, Phys. Rev. D {bf 11} (1975) 3026.
\bibitem{bsG-parameter} Al. Zamolodchikov in 5th Bologna Workshop (2001).
on CFT and Integrable Models, September 26-29, 2001;
Z.~Bajnok, L.~Palla and G.~Tak\'acs, Nucl.~Phys. 
{\bf B622} (2002) 565.
\bibitem{bsG-massless}
H.~Saleur, S.~Skorik, N.~P.~Warner,
Nucl.\ Phys.\ {\bf B441} (1995) 421.
\bibitem{b-state}
S.~Skorik and H.~Saleur, J. Phys. {\bf A28} (1995) 6605; 
P.~Mattison and P.~Dorey, J. Phy. {\bf A33} (2000) 9065;
Z.~Bajnok, L.~Palla and G.~Tak\'acs, Nucl.~Phys. 
{\bf B622} (2002) 548.
\bibitem{ising-TBA}
R.~Chatterjee, Nucl.\ Phys.\ {\bf B468} (1996) 439.
\bibitem{TODA1}
E. Corrigan, P. Dorey, R. Rietdijk, and R. Sasaki,
Phys.\ Lett.\ B333 (1994) 83-91.
\bibitem{TODA2}
V. Fateev, ``Normalization Factors, Reflection
Amplitudes and Integrable Systems'', {\tt hep-th/0103014};
Mod.\ Phys.\ Lett.\ {\bf A16} (2001) 1201;
V.~A.~Fateev and E.~Onofri, 
Nucl.\ Phys.\ {\bf B634} (2002) 546.
\bibitem{TODA3}
C.~Ahn, C.~Kim, C.~Rim, Nucl.\ Phys.\ B628 (2002) 486. 


\end{thebibliography}
\end{document}